\documentclass[a4paper,11pt]{article}
\pdfoutput=1
\usepackage{jheppub}
\usepackage{amsmath,amssymb,mathtools,booktabs,array}
\usepackage{slashed}
\usepackage[section]{placeins}
\usepackage{tikz}
\usetikzlibrary{decorations.pathmorphing,decorations.markings,arrows.meta,calc,shapes.misc,shapes.geometric}

\newcommand{\eb}{\bar\epsilon}
\newcommand{\K}{\mathcal K}
\newcommand{\D}{\mathcal D}
\newcommand{\Fc}{\mathcal F}

\newcommand{\tr}{\operatorname{tr}}
\newcommand{\dd}{\mathrm d}
\newcommand{\MSbar}{\overline{\rm MS}}
\newcommand{\vev}[1]{\langle #1\rangle}
\newcommand{\Lag}{\mathcal L}
\newcommand{\UV}{\Big|_{\rm UV}}

\tikzset{
  gluon/.style={decorate,decoration={coil,aspect=0.75,segment length=3.6pt,amplitude=2.0pt},line width=0.45pt},
  aux/.style={double,double distance=1.5pt,line width=0.45pt},
  quark/.style={line width=0.6pt,postaction={decorate},decoration={markings,mark=at position 0.56 with {\arrow{Stealth[length=2.0mm]}}}},
  heavy/.style={line width=1.4pt,postaction={decorate},decoration={markings,mark=at position 0.56 with {\arrow{Stealth[length=2.6mm]}}}},
  plain/.style={line width=0.6pt},
  ghost/.style={dotted,line width=0.8pt},
  vtx/.style={circle,fill,inner sep=0pt,minimum size=3.6pt},
  mag/.style={rectangle,fill,inner sep=0pt,minimum size=4.6pt},
  gfv/.style={circle,draw,fill=white,inner sep=0pt,minimum size=5pt,line width=0.6pt},
  blob/.style={circle,draw,fill=gray!30,inner sep=0pt,minimum size=9pt,line width=0.6pt},
  bg/.style={circle,draw,fill=white,inner sep=0pt,minimum size=7pt,line width=0.55pt,
     path picture={\draw[line width=0.5pt] (path picture bounding box.south west)--(path picture bounding box.north east)
                   (path picture bounding box.north west)--(path picture bounding box.south east);}},
  ctx/.style={circle,draw,fill=white,inner sep=0pt,minimum size=8pt,line width=0.6pt,
     path picture={\draw[line width=0.6pt] (path picture bounding box.south west)--(path picture bounding box.north east)
                   (path picture bounding box.north west)--(path picture bounding box.south east);}},
  mom/.style={-{Stealth[length=1.6mm]},line width=0.45pt},
  f0v/.style={diamond,draw,fill=white,inner sep=0pt,minimum size=6pt,line width=0.6pt}
}
\newcommand{\mixedline}[2]{\draw[aux] (#1) -- ($(#1)!0.5!(#2)$); \draw[gluon] ($(#1)!0.5!(#2)$) -- (#2);}

\title{Quark-Loop Corrections and Auxiliary Field-Strength\\ Mixing in First-Order QCD}

\author{Hasnat Mohammad Nayeem}

\affiliation{Department of Theoretical Physics, University of Dhaka,\\ Dhaka 1000, Bangladesh}

\emailAdd{hmdnayeem2001@gmail.com}

\abstract{In the first-order formulation of Yang--Mills theory the gluon field strength is an independent auxiliary field $F$, and the only gluon self-interaction is a momentum-independent cubic vertex. We determine how quarks, which couple only to the gauge field, act on this structure. Their divergences are multiples of $f(A)\cdot f(A)$, a term absent from the first-order Lagrangian that no rescaling can produce. Working with the background-field method in a general covariant gauge, we find that the quark sector is renormalized entirely by the mixing of the auxiliary field with the field strength known from pure Yang--Mills theory, $F_0=F+z\,f(A)$, with quark coefficient $z=-\alpha_sT_FN_f/(6\pi\eb)$, together with a rescaling of the gluon field and no rescaling of $F$, so that the renormalized auxiliary field runs into the gluon field strength. When a heavy quark of mass $M$ is integrated out, the same mixing survives as a finite threshold coefficient $c=(\alpha_sT_F/6\pi)\ln(M^2/\mu^2)$, whose scale dependence is exactly the heavy-quark share of the running, and the invariant combination of the couplings reproduces the decoupling of $\alpha_s$. Below the threshold, when the leading power correction is kept as a gauge-sector operator, the auxiliary field is no longer the field strength: $F=f(A)+(g\alpha_sT_F/15\pi M^2)(D_\alpha J_\beta-D_\beta J_\alpha)$, with $J$ the light-quark colour current. We verify this identity against the full theory and show that it separates the long-distance gluon exchange, carried by $F$, from the short-distance heavy-quark effect, carried by a four-quark contact term.}

\begin{document}
\hypersetup{pdftitle={Quark-Loop Corrections and Auxiliary Field-Strength Mixing in First-Order QCD},pdfauthor={Hasnat Mohammad Nayeem}}
\maketitle
\flushbottom

\section{Introduction}
\label{sec:intro}

In the first-order formulation of Yang--Mills theory an antisymmetric tensor field $F^a_{\mu\nu}$ is introduced as an independent variable, and the classical Lagrangian becomes linear in the field strength $f^a_{\mu\nu}(A)$ of the gauge field \cite{Okubo:1979,Halpern:1977,McKeon:1994,Martellini:1997,Cattaneo:1998,Andrasi:2009}. Eliminating $F$ by its algebraic equation of motion gives back the usual theory, so nothing changes classically. Perturbatively, however, the theory looks quite different. The only self-interaction of the gauge field is a cubic vertex $F\!AA$ that does not depend on momenta, the propagator becomes a matrix in the space of $(A,F)$, and every loop graph of the second-order theory is redistributed over graphs with gauge, auxiliary and mixed propagators. This reorganization has been used to simplify perturbative calculations \cite{Brandt:2015,Brandt:2021}, its quantum equivalence with the second-order theory has been analysed through identities between Green's functions with external $F$ and with external field strengths \cite{McKeon:2020}, and it has been combined with the background-field method \cite{DeWitt:1967,tHooft:1973,KlubergStern:1975,Abbott:1981,Abbott:1982,Ichinose:1982,Abbott:1983}, in which the coupling renormalization follows from a single background two-point function through $Z_gZ_B^{1/2}=1$ \cite{McKeon:1994,Brandt:2019bg}. A recurring lesson of this work is that the renormalization of a first-order theory involves not only rescalings of the fields but also mixings between them. In particular, Frenkel and Taylor showed that the BRST renormalization of first-order Yang--Mills theory requires, besides rescalings, a non-linear mixing of the auxiliary field with the full field strength, $F_0=Z_F^{1/2}F+Z_{FA}\,f(A)+\dots$ \cite{Frenkel:2017}, and the same structure controls the renormalization in a background field \cite{Brandt:2019bg}. Related BRST analyses of the background gauge and of a diagonal first-order formulation are given in Refs.~\cite{Frenkel:2018,Brandt:2018diag}.

These studies concern pure Yang--Mills theory. The one-loop $\beta$ function of first-order Yang--Mills theory with massless fermions has been obtained in the Batalin--Vilkovisky framework \cite{Elliott:2018}, but the question of what quarks do to the first-order structure itself has, to our knowledge, not been addressed. The question is natural because quarks couple only to the gauge field and never to the auxiliary field. Every quark loop is therefore a functional of $A$ alone, and its divergence is a multiple of $f(A)\cdot f(A)$. This is precisely the second-order kinetic term that the first-order formulation was built to avoid, and it cannot be produced by any rescaling of $F$, $A$ or $g$; it has to be absorbed by the mixing of Ref.~\cite{Frenkel:2017}, and the question is how. The same issue reappears in finite form at a heavy-quark threshold. Integrating out a heavy quark adds to the light Lagrangian a finite multiple of $f(A)\cdot f(A)$, so the low-energy theory is not of first-order form either, and the momentum-independent vertex structure is lost unless something restores it.

In this paper we resolve both problems, compute the quantities that do so, and determine what happens to the auxiliary field below the threshold. Our first result is the quark contribution to this mixing. The quark sector of first-order QCD is renormalized entirely by a mixing of the auxiliary field with the gluon field strength, $F_0=F+z\,f(A)$, the quark part of the coefficient $Z_{FA}$ of Ref.~\cite{Frenkel:2017}, together with a rescaling of the gluon field equal to the quark part of $Z_B$ and no rescaling of $F$. Unlike the gluon loops, which in general rescale the auxiliary field as well as mixing it, the quarks contribute to the mixing only. We compute $z=-\alpha_sT_FN_f/(6\pi\eb)$, show that it is independent of the gauge parameter and of the quark masses, show that it preserves the background value $f(B)$ of the auxiliary field, and derive the resulting renormalization-group equation, in which the renormalized auxiliary field runs into $f(A)$ at the rate $\alpha_sT_FN_f/3\pi$. Our second result is that heavy-quark decoupling acts on the auxiliary field through the same mechanism. After a heavy quark is integrated out, the first-order form is restored by a finite mixing with coefficient $c=(\alpha_sT_F/6\pi)\ln(M^2/\mu^2)$, followed by the rescaling that encodes the decoupling of $\alpha_s$ \cite{Bernreuther:1982,Larin:1995,Chetyrkin:1998}. The scale dependence of $c$ is exactly the heavy flavour's share of the running found in the first result, and the ultraviolet and threshold mixings together equal $-\frac12$ times the zero-momentum heavy-quark loop. At order $1/M^2$ the heavy-quark power corrections \cite{Novikov:1984,Cho:1995} can be written with momentum-independent vertices only, so the defining simplification of the first-order formalism survives decoupling. Our third result concerns the auxiliary field below the threshold. The algebraic identity $F=f(A)$, which underlies the structural identities of the first-order formalism \cite{McKeon:2020}, acquires a contact term, $F=f(A)+(g\alpha_sT_F/15\pi M^2)(D_\alpha J_\beta-D_\beta J_\alpha)$, proportional to the covariant curl of the light-quark colour current, in the form of the light Lagrangian in which the leading power correction is kept as a gauge-sector operator. We derive it in two ways, show that its divergence is the non-Abelian Uehling-corrected field equation, and check it against the full theory. The check shows what the identity means: in the light theory the field strength $f(A)$ carries the heavy-quark contact term, as the full theory requires, while the auxiliary field is a pure gluon pole, so the first-order variables separate long-distance exchange from the short-distance effect of the heavy quark.

These results do not follow from the equivalence of the first-order and second-order theories, which fixes only the invariant combination of the gauge couplings, nor from the pure Yang--Mills analyses of Refs.~\cite{Brandt:2019bg,Frenkel:2017}, which contain no quarks. They concern the auxiliary field itself, and they are needed in any first-order calculation with quarks beyond one loop, where $z$ is the counterterm of every quark-loop subdivergence, and in any first-order effective theory below a heavy-quark threshold.

Section~\ref{sec:setup} sets up first-order background-field QCD and the three couplings of its gauge sector. Section~\ref{sec:mixing} contains the quark loop and the first result, and Section~\ref{sec:threshold} the second and third. We conclude in Section~\ref{sec:concl}. The appendices contain the supporting one-loop calculations in first-order variables: the conventions and integrals (Appendix~\ref{app:conv}), the Feynman rules and the graphs that vanish (Appendix~\ref{app:rules}), the quark self-energy and the background quark--gluon vertex, which verify the Ward identity used in Section~\ref{sec:mixing} (Appendix~\ref{sec:quark}), the gluonic part of the background two-point function, which completes $Z_B$ (Appendix~\ref{sec:twopoint}), and the stationarity argument used in Section~\ref{sec:threshold} (Appendix~\ref{app:stationarity}).

Throughout we work in Minkowski space with $\eta_{\mu\nu}={\rm diag}(+1,-1,-1,-1)$ and in dimensional regularization with $d=4-2\epsilon$, and we define
\begin{equation}
\frac1\eb\equiv\frac1\epsilon-\gamma_E+\ln4\pi,\qquad \K\equiv\frac{g^2}{16\pi^2\eb},
\label{eq:K}
\end{equation}
so that subtraction of the poles in $\eb$ defines the $\MSbar$ scheme \cite{tHooft:1972,Bardeen:1978}; for the standard second-order results used for comparison we follow Refs.~\cite{Peskin:1995,Pascual:1984,Muta:2010,Grozin:2007,Collins:1984}. The generators of $SU(N)$ in the fundamental representation satisfy $[T^a,T^b]=if^{abc}T^c$, ${\rm Tr}(T^aT^b)=T_F\delta^{ab}$, $T^aT^a=C_F\mathbf 1$ and $f^{acd}f^{bcd}=C_A\delta^{ab}$. We write $(X\times Y)^a\equiv f^{abc}X^bY^c$ and $X\cdot Y\equiv X^a_{\mu\nu}Y^{a\mu\nu}$, summed over all $\mu,\nu$, for antisymmetric tensors.

\section{First-order background-field QCD}
\label{sec:setup}

The first-order Lagrangian of QCD with quarks of mass $m$ is
\begin{equation}
\Lag=\frac14F\cdot F-\frac12F\cdot f(A)+\bar\psi\left(i\gamma^\mu D_\mu-m\right)\psi,
\label{eq:L}
\end{equation}
with
\begin{equation}
f^a_{\mu\nu}(A)=\partial_\mu A^a_\nu-\partial_\nu A^a_\mu+gf^{abc}A^b_\mu A^c_\nu,\qquad D_\mu\psi=\left(\partial_\mu-igA^a_\mu T^a\right)\psi .
\label{eq:f}
\end{equation}
The field equation of $F$ is algebraic, $F=f(A)$, and substituting it gives $-\frac14f\cdot f$; since the $F$-dependence is Gaussian the substitution is exact in the path integral \cite{Halpern:1977,McKeon:2020}. The field equation of $A$ is
\begin{equation}
(D_\mu F^{\mu\nu})^a=-g\,J^{a\nu},\qquad J^{a\nu}=\bar\psi\gamma^\nu T^a\psi ,
\label{eq:eomA}
\end{equation}
an exact relation of the full theory that involves only the elementary field $F$.

To set up the background-field method we write
\begin{equation}
A^a_\mu=B^a_\mu+Q^a_\mu,\qquad F^a_{\mu\nu}=f^a_{\mu\nu}(B)+G^a_{\mu\nu},
\label{eq:split}
\end{equation}
with background $B$ and quantum fields $Q$ and $G$. The background value of $F$ is chosen equal to its classical value $f(B)$, which removes any term linear in $G$ (another choice $\bar F$ would leave $\frac12G\cdot(\bar F-f(B))$ and auxiliary-field tadpoles). With $(\D_\mu X)^a=\partial_\mu X^a+gf^{abc}B^b_\mu X^c$ and $\Fc(Q)_{\mu\nu}=\D_\mu Q_\nu-\D_\nu Q_\mu$, we have $f(B+Q)=f(B)+\Fc(Q)+g\,Q\times Q$, and dropping the term linear in quantum fields \cite{Abbott:1981},
\begin{equation}
\Lag_{\rm YM}=-\frac14f(B)\cdot f(B)+\frac14G\cdot G-\frac12G\cdot\Fc(Q)-\frac g2\left[f(B)+G\right]\cdot(Q\times Q).
\label{eq:LYM}
\end{equation}
The term $-\frac g2f(B)\cdot(Q\times Q)$ is the magnetic coupling of the quantum gluon to the background, and $-\frac g2G\cdot(Q\times Q)$ is the momentum-independent cubic vertex of the first-order formalism. In the background-covariant gauge,
\begin{equation}
\begin{gathered}
\Lag_{\rm gf}+\Lag_{\rm gh}=-\frac1{2\xi}\left[(\D^\mu Q_\mu)^a\right]^2-\bar c^a\left[\D^\mu D_\mu(B+Q)\right]^{ab}c^b,\\
\Lag_q=\bar\psi\left(i\gamma^\mu D_\mu(B)-m\right)\psi+g\bar\psi\gamma^\mu T^a\psi\,Q^a_\mu .
\end{gathered}
\label{eq:gf}
\end{equation}
The total Lagrangian is invariant under background gauge transformations, under which $B$ transforms as a gauge field and $Q$, $G$, $f(B)$, $F$ and the ghosts transform in the adjoint representation.

Completing the square in the part of $\Lag_{\rm YM}+\Lag_{\rm gf}$ quadratic in the quantum fields gives
\begin{equation}
\Lag^{(2)}=\frac14\big(G-\Fc(Q)\big)^2-\frac14\Fc(Q)\cdot\Fc(Q)-\frac g2f(B)\cdot(Q\times Q)-\frac1{2\xi}(\D\cdot Q)^2 .
\label{eq:L2sq}
\end{equation}
The shifted field $G'=G-\Fc(Q)$ decouples with unit Jacobian, and the remaining terms are the quadratic Lagrangian of the second-order background-field method \cite{Abbott:1981}. Since quarks and ghosts do not couple to $G$,
\begin{equation}
\Gamma^{(1)}_{\rm first\ order}[B,\psi,\bar\psi]=\Gamma^{(1)}_{\rm second\ order}[B,\psi,\bar\psi]+{\rm const}.
\label{eq:equiv}
\end{equation}
At $B=0$ the same algebra gives the propagators. The gluon propagator is the second-order one, and, since $G=G'+f_0(Q)$ with $f_0(Q)=\partial\wedge Q$ and $\vev{G'G'}=2iI$, $I_{\mu\nu,\rho\sigma}=\frac12(\eta_{\mu\rho}\eta_{\nu\sigma}-\eta_{\mu\sigma}\eta_{\nu\rho})$,
\begin{align}
\vev{Q^a_\mu Q^b_\nu}(p)&=\delta^{ab}D_{\mu\nu}(p)=\frac{-i\delta^{ab}}{p^2+i0}\left[\eta_{\mu\nu}-(1-\xi)\frac{p_\mu p_\nu}{p^2}\right],\label{eq:DQQ}\\
\vev{G^a_{\mu\nu}Q^b_\rho}(p)&=-\delta^{ab}\,\frac{p_\mu\eta_{\nu\rho}-p_\nu\eta_{\mu\rho}}{p^2+i0},\label{eq:DGQ}\\
\vev{G^a_{\mu\nu}G^b_{\rho\sigma}}(p)&=2i\delta^{ab}\left[I_{\mu\nu,\rho\sigma}-\Pi^L_{\mu\nu,\rho\sigma}(p)\right],\label{eq:DGG}
\end{align}
with $\Pi^L$ the longitudinal projector on antisymmetric tensors. The auxiliary propagators are independent of $\xi$. Their structure is shown in Fig.~\ref{fig:propdecomp}; the ghost and quark propagators are standard, and all Feynman rules are collected in Appendix~\ref{app:rules}.

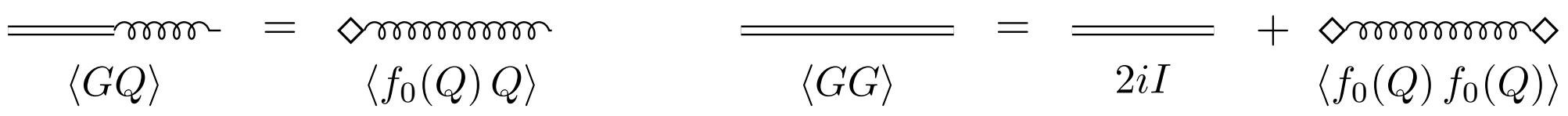
\begin{figure}[t]
\centering
\begin{tikzpicture}
  \begin{scope}
    \draw[aux] (0,0)--(0.9,0); \draw[gluon] (0.9,0)--(1.8,0);
    \node at (2.3,0) {$=$};
    \node[f0v] (d) at (2.9,0) {}; \draw[gluon] (d)--(4.6,0);
    \node[below] at (0.9,-0.15) {\small $\vev{GQ}$};
    \node[below] at (3.75,-0.15) {\small $\vev{f_0(Q)\,Q}$};
  \end{scope}
  \begin{scope}[shift={(6.2,0)}]
    \draw[aux] (0,0)--(1.8,0);
    \node at (2.3,0) {$=$};
    \draw[aux] (2.8,0)--(4.0,0); \node[below] at (3.4,-0.15) {\small $2iI$};
    \node at (4.5,0) {$+$};
    \node[f0v] (d1) at (5.0,0) {}; \node[f0v] (d2) at (6.8,0) {};
    \draw[gluon] (d1)--(d2);
    \node[below] at (0.9,-0.15) {\small $\vev{GG}$};
    \node[below] at (5.9,-0.15) {\small $\vev{f_0(Q)\,f_0(Q)}$};
  \end{scope}
\end{tikzpicture}
\caption{Propagators involving the auxiliary field (double line). The diamond is an insertion of the Abelian field strength $f_0(Q)=\partial\wedge Q$. The mixed propagator is a gluon line ending on such an insertion; the auxiliary propagator is the local term $2iI$ plus a gluon line with insertions at both ends.}
\label{fig:propdecomp}
\end{figure}

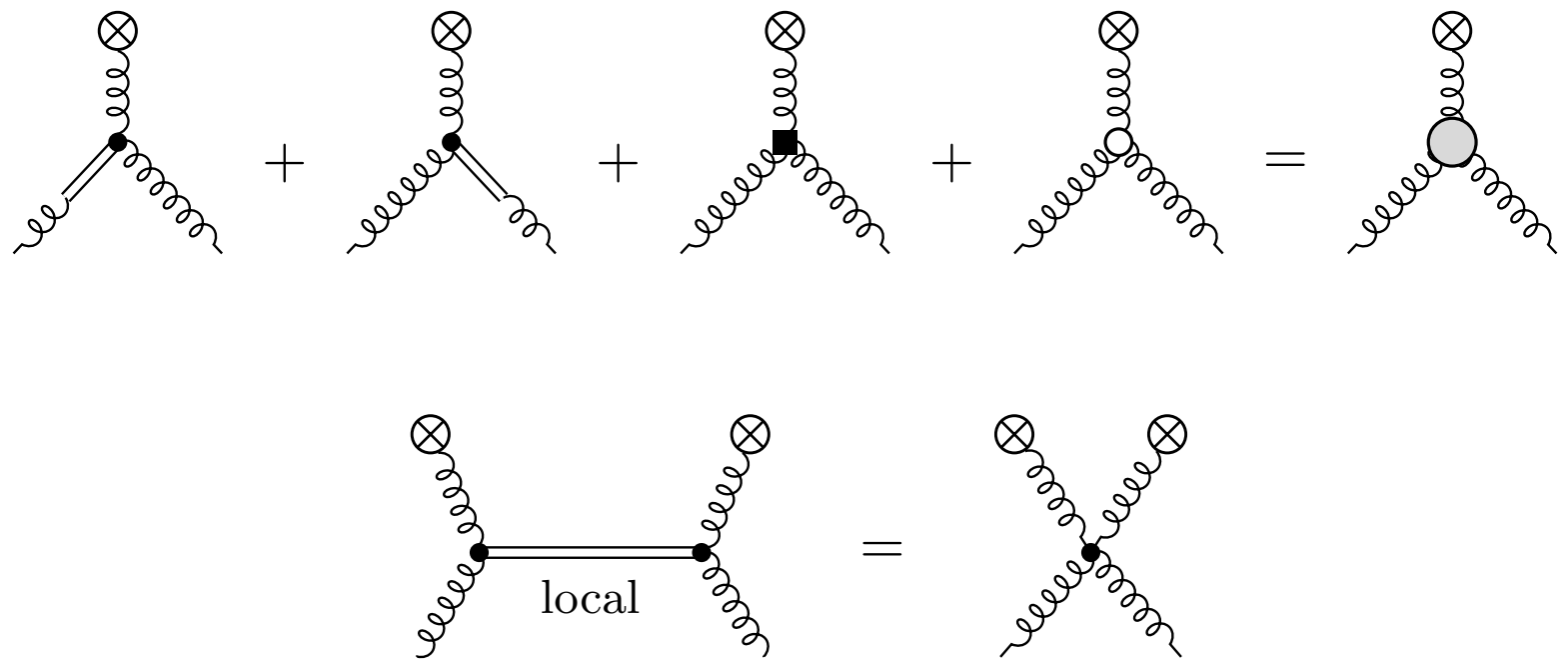
\begin{figure}[t]
\centering
\begin{tikzpicture}[scale=0.92]
 \newcommand{\bgthree}[3]{%
   \begin{scope}[shift={(#1,0)}]
     \node[bg] (B) at (0,1.15) {};
     \draw[gluon] (B) -- (0,0.35);
     \coordinate (V) at (0,0.35);
     #2
     #3
   \end{scope}}
 \bgthree{0}{\coordinate (L) at (-0.75,-0.45); \coordinate (R) at (0.75,-0.45); \mixedline{V}{L}; \draw[gluon] (V)--(R);}{\node[vtx] at (V) {};}
 \node at (1.2,0.2) {$+$};
 \bgthree{2.4}{\coordinate (L) at (-0.75,-0.45); \coordinate (R) at (0.75,-0.45); \draw[gluon] (V)--(L); \mixedline{V}{R};}{\node[vtx] at (V) {};}
 \node at (3.6,0.2) {$+$};
 \bgthree{4.8}{\coordinate (L) at (-0.75,-0.45); \coordinate (R) at (0.75,-0.45); \draw[gluon] (V)--(L); \draw[gluon] (V)--(R);}{\node[mag] at (V) {};}
 \node at (6.0,0.2) {$+$};
 \bgthree{7.2}{\coordinate (L) at (-0.75,-0.45); \coordinate (R) at (0.75,-0.45); \draw[gluon] (V)--(L); \draw[gluon] (V)--(R);}{\node[gfv] at (V) {};}
 \node at (8.4,0.2) {$=$};
 \bgthree{9.6}{\coordinate (L) at (-0.75,-0.45); \coordinate (R) at (0.75,-0.45); \draw[gluon] (V)--(L); \draw[gluon] (V)--(R);}{\node[blob] at (V) {};}
 \begin{scope}[shift={(2.6,-2.6)}]
   \coordinate (V1) at (0,0); \coordinate (V2) at (1.6,0);
   \node[bg] (B1) at (-0.35,0.85) {}; \node[bg] (B2) at (1.95,0.85) {};
   \draw[gluon] (B1)--(V1); \draw[gluon] (B2)--(V2);
   \draw[aux] (V1)--(V2);
   \draw[gluon] (V1)--(-0.45,-0.75); \draw[gluon] (V2)--(2.05,-0.75);
   \node[vtx] at (V1) {}; \node[vtx] at (V2) {};
   \node[below] at (0.8,-0.05) {\footnotesize local};
   \node at (2.9,0) {$=$};
   \coordinate (W) at (4.4,0);
   \node[bg] (C1) at (3.85,0.85) {}; \node[bg] (C2) at (4.95,0.85) {};
   \draw[gluon] (C1)--(W); \draw[gluon] (C2)--(W);
   \draw[gluon] (W)--(3.75,-0.75); \draw[gluon] (W)--(5.05,-0.75);
   \node[vtx] at (W) {};
 \end{scope}
\end{tikzpicture}
\caption{Diagrammatic dictionary between the first-order and second-order background-field methods. Upper row: the GBQ vertex with its auxiliary leg (double line) closed on a mixed propagator, in both assignments, plus the magnetic vertex (square) and the gauge-fixing vertex (open circle), equals the second-order background vertex \eqref{eq:Abbottvertex} (blob). Lower row: two GBQ vertices joined by the local part of the auxiliary propagator reproduce the second-order seagull vertex. Crossed circles denote background fields.}
\label{fig:dictionary}
\end{figure}

Turning to the vertices, both quark--gluon vertices are $ig\gamma_\mu T^a$. With all momenta incoming, a background field of momentum $k$ and gluons of momenta $p$ and $q$, the background vertices with two quantum legs are
\begin{equation}
\begin{gathered}
\big[G^a_{\mu\nu}B^b_\lambda Q^c_\rho\big]=-igf^{abc}\eta_{\mu\lambda}\eta_{\nu\rho}\Big|_{[\mu\nu]},\\
\big[B^a_\lambda Q^b_\rho Q^c_\sigma\big]_{\rm mag}=-gf^{abc}\left(k_\rho\eta_{\lambda\sigma}-k_\sigma\eta_{\lambda\rho}\right),\qquad
\big[B^a_\lambda Q^b_\rho Q^c_\sigma\big]_{\rm gf}=\frac g\xi f^{abc}\left(p_\rho\eta_{\lambda\sigma}-q_\sigma\eta_{\lambda\rho}\right),
\end{gathered}
\label{eq:Vertices}
\end{equation}
the background--auxiliary--gluon vertex (GBQ), the magnetic vertex and the gauge-fixing vertex. Equation~\eqref{eq:DGQ} shows that the mixed propagator is the propagator of $f_0(Q)$. A GBQ vertex whose auxiliary leg is closed on a mixed propagator therefore acts as the vertex $gf^{abc}(p_\lambda\eta_{\sigma\rho}-p_\sigma\eta_{\lambda\rho})$ followed by an ordinary gluon line. Adding both assignments of the auxiliary leg, the magnetic and the gauge-fixing vertex gives
\begin{equation}
\Gamma^{abc}_{\lambda\rho\sigma}(k,p,q)=gf^{abc}\left[\eta_{\rho\sigma}(p-q)_\lambda+\eta_{\lambda\sigma}\Big(q-k+\frac p\xi\Big)_\rho+\eta_{\lambda\rho}\Big(k-p-\frac q\xi\Big)_\sigma\right],
\label{eq:Abbottvertex}
\end{equation}
the background vertex of the second-order method \cite{Abbott:1981}, and two GBQ vertices joined by the local part $2iI$ of Eq.~\eqref{eq:DGG} give the second-order seagull vertex. This dictionary, Fig.~\ref{fig:dictionary}, shows how first-order graphs reassemble into second-order ones (Appendices~\ref{sec:quark} and \ref{sec:twopoint}). It also shows that the first-order computation is not merely a rewriting: one second-order graph splits into several first-order graphs with different propagators and different gauge dependence.

The analysis below rests on a simple observation. Consider the general gauge sector
\begin{equation}
\Lag_{\kappa\lambda\tau}=\frac\kappa4F\cdot F-\frac\lambda2F\cdot f(A)-\frac\tau4f(A)\cdot f(A),
\label{eq:Lklt}
\end{equation}
of which Eq.~\eqref{eq:L} is the case $(1,1,0)$. Eliminating $F=(\lambda/\kappa)f$ gives
\begin{equation}
\Lag_{\kappa\lambda\tau}\Big|_{F=\lambda f/\kappa}=-\frac{\mathcal Z}4\,f\cdot f,\qquad \mathcal Z\equiv\frac{\lambda^2}\kappa+\tau,
\label{eq:Zphys}
\end{equation}
so only $\mathcal Z$ is physical. Three field redefinitions with field-independent Jacobians act on the couplings. A rescaling $F\to\zeta F$ gives $(\zeta^2\kappa,\zeta\lambda,\tau)$. A gluon rescaling at fixed $gA$, $A=sA'$ and $g=g'/s$, gives $(\kappa,s\lambda,s^2\tau)$. A mixing $F\to F+c\,f(A)$ gives
\begin{equation}
(\kappa,\lambda,\tau)\ \to\ \left(\kappa,\;\lambda-c\kappa,\;\tau+2c\lambda-c^2\kappa\right).
\label{eq:mixmap}
\end{equation}
The first and third leave $\mathcal Z$ invariant; the gluon rescaling multiplies it by $s^2$ and is the physical renormalization of the coupling. The key point is that neither rescaling can turn $\tau=0$ into $\tau\neq0$ or back: an $f\cdot f$ term can be produced or removed only by the mixing \eqref{eq:mixmap}. All three act on the full fields $A=B+Q$ and $F=f(B)+G$, which transform covariantly, so they respect background gauge invariance.

For the renormalization we write $B_0=Z_B^{1/2}B$, $g_0=\mu^\epsilon Z_gg$, $\psi_0=Z_\psi^{1/2}\psi$ and $m_0=Z_mm$. Background gauge invariance requires the background covariant derivative to be renormalized as a whole, so \cite{Abbott:1981}
\begin{equation}
Z_gZ_B^{1/2}=1,
\label{eq:ZgZB}
\end{equation}
and the counterterms needed at one loop are
\begin{equation}
\begin{split}
\delta\Lag={}&-\frac{\delta Z_B}4f(B)\cdot f(B)+\delta Z_\psi\,\bar\psi i\slashed\partial\psi-(\delta Z_\psi+\delta Z_m)\,m\bar\psi\psi+\delta Z_{1B}\,g\bar\psi\slashed B\psi,\\
&Z_{1B}\equiv Z_gZ_\psi Z_B^{1/2}.
\end{split}
\label{eq:dL}
\end{equation}
Equation~\eqref{eq:ZgZB} implies $Z_{1B}=Z_\psi$, which we verify in first-order variables in Appendix~\ref{sec:quark}.

\section{Quark loops renormalize the auxiliary field by mixing}
\label{sec:mixing}

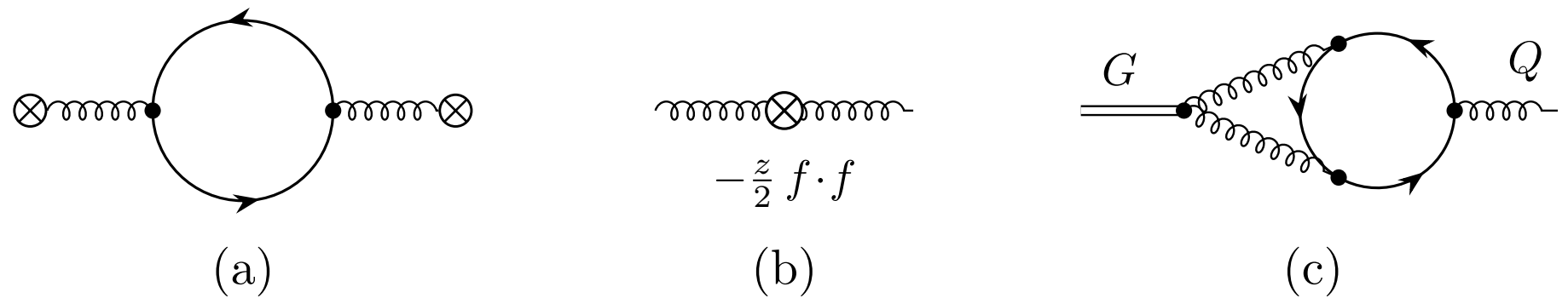
\begin{figure}[t]
\centering
\begin{tikzpicture}
 \begin{scope}
  \node[bg] (B1) at (-0.95,0) {}; \node[bg] (B2) at (2.35,0) {};
  \coordinate (v1) at (0,0); \coordinate (v2) at (1.4,0);
  \draw[gluon] (B1)--(v1); \draw[gluon] (v2)--(B2);
  \draw[quark] (v2) arc[start angle=0,end angle=180,radius=0.7];
  \draw[quark] (v1) arc[start angle=180,end angle=360,radius=0.7];
  \node[vtx] at (v1) {}; \node[vtx] at (v2) {};
  \node at (0.7,-1.25) {(a)};
 \end{scope}
 \begin{scope}[shift={(3.9,0)}]
  \coordinate (X) at (1.0,0);
  \draw[gluon] (0,0)--(X); \draw[gluon] (X)--(2.0,0);
  \node[ctx] at (X) {};
  \node[below] at (1.0,-0.25) {\small $-\frac z2\,f\!\cdot\! f$};
  \node at (1.0,-1.25) {(b)};
 \end{scope}
 \begin{scope}[shift={(7.2,0)}]
  \coordinate (G) at (0,0); \coordinate (V) at (0.8,0); \coordinate (C) at (2.3,0);
  \draw[aux] (G)--(V);
  \coordinate (P1) at ($(C)+(120:0.6)$); \coordinate (P2) at ($(C)+(240:0.6)$); \coordinate (P3) at ($(C)+(0:0.6)$);
  \draw[gluon] (V)--(P1); \draw[gluon] (V)--(P2);
  \draw[quark] (P3) arc[start angle=0,end angle=120,radius=0.6];
  \draw[quark] (P1) arc[start angle=120,end angle=240,radius=0.6];
  \draw[quark] (P2) arc[start angle=240,end angle=360,radius=0.6];
  \draw[gluon] (P3)--(3.7,0);
  \node[vtx] at (V) {}; \node[vtx] at (P1) {}; \node[vtx] at (P2) {}; \node[vtx] at (P3) {};
  \node at (0.3,0.32) {\small $G$}; \node at (3.45,0.38) {\small $Q$};
  \node at (1.8,-1.25) {(c)};
 \end{scope}
\end{tikzpicture}
\caption{(a) The quark loop with two gluon legs (shown with background legs; the same graph with quantum legs enters $\vev{QQ}$). (b) The counterterm insertion $-\frac z2f(A)\cdot f(A)$ (crossed circle) on a gluon line, generated by the mixing \eqref{eq:Fbare}. (c) The lowest graph in which a quark loop reaches the auxiliary field; it needs the cubic vertex $GQQ$ and has two loops.}
\label{fig:quarksector}
\end{figure}

We first compute the quark loop with two gluon legs, Fig.~\ref{fig:quarksector}(a). The legs may be background or quantum gluons; the vertices are the same. With $N_f$ flavours of common mass $m$,
\begin{equation}
i\Pi^{ab}_{\mu\nu}(k)\Big|_q=-g^2T_FN_f\,\delta^{ab}\mu^{2\epsilon}\!\int\!\frac{\dd^dp}{(2\pi)^d}\,\frac{N_{\mu\nu}}{(p^2-m^2)\left[(p+k)^2-m^2\right]},
\label{eq:qloop}
\end{equation}
where the factor $-(ig)^2\,i^2\,{\rm Tr}(T^aT^b)$ includes the minus sign of the fermion loop, and
\begin{equation}
\begin{split}
N_{\mu\nu}&=\tr\left[\gamma_\mu(\slashed p+m)\gamma_\nu(\slashed p+\slashed k+m)\right]\\
&=4\left[p_\mu(p+k)_\nu+p_\nu(p+k)_\mu-\eta_{\mu\nu}\left(p\cdot(p+k)-m^2\right)\right].
\end{split}
\label{eq:Ntrace}
\end{equation}
Transversality holds before any integral is done. Contracting with $k^\mu$ and writing $\slashed k=(\slashed p+\slashed k-m)-(\slashed p-m)$, the integrand becomes the difference of two tadpoles related by a shift of the loop momentum, so $k^\mu\Pi^{ab}_{\mu\nu}=0$ in dimensional regularization. Explicitly, we combine the denominators as $(1-x)(p^2-m^2)+x[(p+k)^2-m^2]=\ell^2-\Delta_q$, with $\ell=p+xk$ and $\Delta_q=m^2-x(1-x)k^2$. Dropping terms odd in $\ell$ and using $\ell_\mu\ell_\nu\to\ell^2\eta_{\mu\nu}/d$,
\begin{equation}
N_{\mu\nu}\to4\left[\Big(\frac2d-1\Big)\ell^2\eta_{\mu\nu}-2x(1-x)k_\mu k_\nu+\left(x(1-x)k^2+m^2\right)\eta_{\mu\nu}\right].
\label{eq:Nshift}
\end{equation}
The identity $\int\ell^2(\ell^2-\Delta)^{-2}=\frac d{d-2}\Delta\int(\ell^2-\Delta)^{-2}$ turns $(\frac2d-1)\ell^2$ into $-\Delta_q$, and $-\Delta_q+x(1-x)k^2+m^2=2x(1-x)k^2$. With
\begin{equation}
I_2(\Delta)\equiv\mu^{2\epsilon}\!\int\!\frac{\dd^d\ell}{(2\pi)^d}\,\frac1{(\ell^2-\Delta+i0)^2}=\frac{i}{16\pi^2}\left[\frac1\eb-\ln\frac{\Delta}{\mu^2}\right]+O(\epsilon),
\label{eq:I2}
\end{equation}
we obtain the exactly transverse result
\begin{equation}
\begin{gathered}
\Pi^{ab}_{\mu\nu}(k)\Big|_q=\delta^{ab}\left(k_\mu k_\nu-k^2\eta_{\mu\nu}\right)\Pi_q(k^2),\\
\Pi_q(k^2)=\frac{g^2T_FN_f}{2\pi^2}\int_0^1\dd x\,x(1-x)\left[\frac1\eb-\ln\frac{m^2-x(1-x)k^2-i0}{\mu^2}\right].
\end{gathered}
\label{eq:Piq}
\end{equation}
Since $\int_0^1x(1-x)\,\dd x=\frac16$, the pole is
\begin{equation}
\Pi_q(k^2)\Big|_{\rm UV}=\frac43T_FN_f\,\K .
\label{eq:Piqpole}
\end{equation}
It is independent of the quark mass because a local polynomial of dimension two that is transverse cannot contain $m^2\eta_{\mu\nu}$. The tree-level two-point vertex from $-\frac14f\cdot f$ is $i\delta^{ab}(k_\mu k_\nu-k^2\eta_{\mu\nu})$, so the loop is the vertex of $-\frac14\Pi_q\,f\cdot f$.

This divergence is not of first-order form. Quarks couple only to $A=B+Q$. Every quark loop, with any number of background and quantum gluon legs, is therefore part of the functional determinant ${\rm Det}(i\slashed D(A)-m)$, which depends on $A$ alone and never on $G$. Its one-loop ultraviolet divergence is local, gauge invariant and of dimension four, so it is a multiple of $f(A)\cdot f(A)$, and Eq.~\eqref{eq:Piqpole} fixes the multiple:
\begin{equation}
\Gamma_q\Big|_{\rm UV}=-\frac14\,a\int f(A)\cdot f(A),\qquad a\equiv\frac43T_FN_f\K .
\label{eq:GammaqUV}
\end{equation}
The counterterm it requires, $+\frac a4f(A)\cdot f(A)$, is a $\tau$ term in the language of Eq.~\eqref{eq:Lklt}. By the observation below Eq.~\eqref{eq:mixmap}, it cannot be produced by multiplicative renormalization of $F$, $A$ and $g$.

The quark divergence must nevertheless be absorbed in a bare Lagrangian of first-order form, $\frac14F_0\cdot F_0-\frac12F_0\cdot f_{g_0}(A_0)$. We therefore allow, in addition to the rescalings, the mixing of $F$ with $f(A)$ that already appears in the BRST renormalization of pure first-order Yang--Mills theory \cite{Frenkel:2017}:
\begin{equation}
F_0=Z_F^{1/2}\big(F+z\,f(A)\big),\qquad A_0=Z_A^{1/2}A,\qquad g_0Z_A^{1/2}=\mu^\epsilon g .
\label{eq:Fbare}
\end{equation}
The last relation keeps $g_0A_0=\mu^\epsilon gA$, so the quark--gluon vertex is not renormalized beyond $Z_\psi$, as required by Eq.~\eqref{eq:ZgZB}; it also gives $f_{g_0}(A_0)=Z_A^{1/2}f_g(A)$. Substituting Eq.~\eqref{eq:Fbare}, the two terms of the bare gauge Lagrangian become
\begin{align}
\frac14F_0\cdot F_0&=\frac{Z_F}4\Big(F\cdot F+2z\,F\cdot f(A)+z^2f(A)\cdot f(A)\Big),\\
-\frac12F_0\cdot f_{g_0}(A_0)&=-\frac12Z_F^{1/2}Z_A^{1/2}\Big(F\cdot f(A)+z\,f(A)\cdot f(A)\Big).
\end{align}
Collecting the coefficients of $F\cdot F$, $F\cdot f$ and $f\cdot f$, and keeping first-order terms in $\delta Z_F$, $\delta Z_A$ and $z$ (so that $z^2$ and $z\,\delta Z$ are dropped),
\begin{equation}
\Lag_0=\frac14(1+\delta Z_F)F\cdot F-\frac12\Big(1+\frac12\delta Z_F+\frac12\delta Z_A-z\Big)F\cdot f(A)-\frac z2f(A)\cdot f(A).
\label{eq:Fbareexp}
\end{equation}
The coefficients follow from the one-particle-irreducible functions. At one loop the quarks do not contribute to $\vev{GG}$ or $\vev{GQ}$, because $G$ couples to quarks only through the cubic vertex $GQQ$ and the first such graph, Fig.~\ref{fig:quarksector}(c), has two loops. The quark parts of the coefficients of $F\cdot F$ and $F\cdot f$ must therefore vanish, and the coefficient of $f\cdot f$ must equal the counterterm $+\frac a4$. This gives $\delta Z_F^{(q)}=0$, $\delta Z_A^{(q)}=2z^{(q)}$ and $-\frac12z^{(q)}=\frac a4$, that is,
\begin{equation}
z^{(q)}=-\frac23T_FN_f\K=-\frac{\alpha_sT_FN_f}{6\pi\eb},\qquad \delta Z_A^{(q)}=-\frac43T_FN_f\K,\qquad \delta Z_F^{(q)}=0 .
\label{eq:zq}
\end{equation}
With these values the coefficients of $F\cdot F$ and $F\cdot f$ in Eq.~\eqref{eq:Fbareexp} keep their tree values, and the only counterterm vertex is $-\frac z2f(A)\cdot f(A)$ on gluon lines, Fig.~\ref{fig:quarksector}(b). It is not a new term in the Lagrangian: in first-order variables it is generated by the mixing in Eq.~\eqref{eq:Fbare}.

Equation~\eqref{eq:zq} has four consequences. First, the gluon field, background and quantum alike, is rescaled by the quark part of $Z_B$; combined with the gluonic part, which we compute from the first-order rules in Appendix~\ref{sec:twopoint} (see also Refs.~\cite{Abbott:1981,Brandt:2019bg}), it gives $Z_B=1+(\frac{11}3C_A-\frac43T_FN_f)\K$, $Z_g=Z_B^{-1/2}$ and the familiar $\beta_0=\frac{11}3C_A-\frac43T_FN_f$ \cite{Gross:1973,Politzer:1973}. Second, the auxiliary field is not rescaled by the quarks at all; their entire effect on it is the mixing $z^{(q)}=\frac12\delta Z_B^{(q)}$. In the notation of Frenkel and Taylor, $F_0=Z_F^{1/2}F+Z_{FA}f(A)$, this is the quark part $Z_{FA}^{(q)}=z^{(q)}$ of the mixing coefficient, with $Z_F^{(q)}=1$; in pure Yang--Mills theory the gluon loops contribute to both $Z_F$ and $Z_{FA}$ \cite{Frenkel:2017}. Third, the split \eqref{eq:split} is stable: the background value of $F_0$ is $(1+z^{(q)})f(B)$, which equals $f(B_0)=Z_B^{1/2}f(B)$ at this order, so the choice $f(B)$ for the auxiliary background survives renormalization. In terms of the quantum fields, Eq.~\eqref{eq:Fbare} reads $G_0=G+z^{(q)}[f(B+Q)-f(B)]$, a mixing of $G$ with $\Fc(Q)+g\,Q\times Q$. Fourth, $z^{(q)}$ is independent of the gauge parameter, because the quark loop contains no gluon propagator, and of the quark masses, by Eq.~\eqref{eq:Piqpole}.

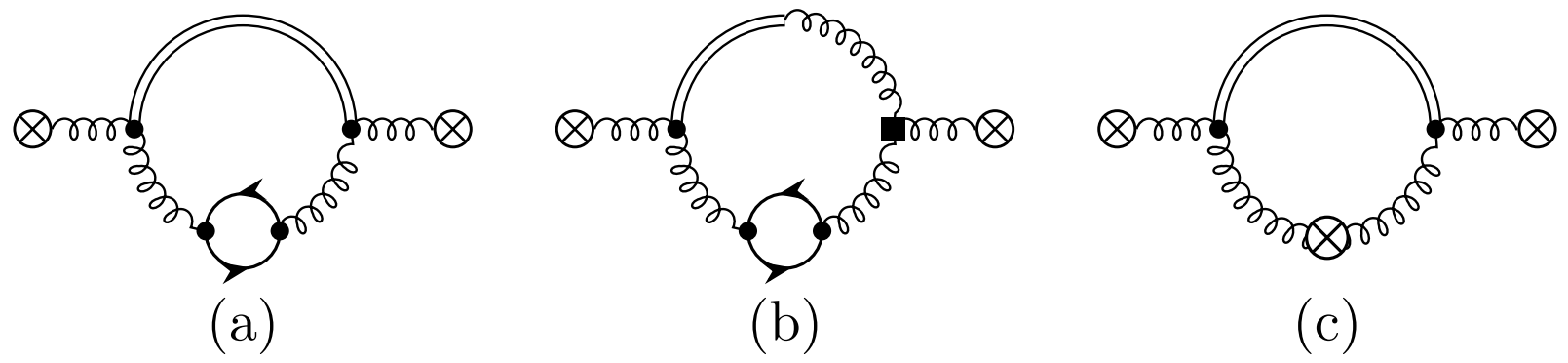
\begin{figure}[t]
\centering
\begin{tikzpicture}[scale=0.92]
\newcommand{\loopframe}{%
  \node[bg] (BL) at (-0.75,0) {}; \node[bg] (BR) at (2.35,0) {};
  \draw[gluon] (BL)--(0,0); \draw[gluon] (1.6,0)--(BR);}
\newcommand{\lowerbubble}{%
  \draw[gluon] (0,0) arc[start angle=180,end angle=250,radius=0.8];
  \draw[gluon] ($(0.8,0)+(290:0.8)$) arc[start angle=290,end angle=360,radius=0.8];
  \draw[quark] ($(0.8,0)+(250:0.8)$) arc[start angle=180,end angle=360,radius=0.274];
  \draw[quark] ($(0.8,0)+(290:0.8)$) arc[start angle=0,end angle=180,radius=0.274];
  \node[vtx] at ($(0.8,0)+(250:0.8)$) {}; \node[vtx] at ($(0.8,0)+(290:0.8)$) {};}
 \begin{scope}
  \loopframe
  \draw[aux] (0,0) arc[start angle=180,end angle=0,radius=0.8];
  \lowerbubble
  \node[vtx] at (0,0) {}; \node[vtx] at (1.6,0) {};
  \node at (0.8,-1.45) {(a)};
 \end{scope}
 \begin{scope}[shift={(4.0,0)}]
  \loopframe
  \draw[aux] (0,0) arc[start angle=180,end angle=90,radius=0.8]; \draw[gluon] (0.8,0.8) arc[start angle=90,end angle=0,radius=0.8];
  \lowerbubble
  \node[vtx] at (0,0) {}; \node[mag] at (1.6,0) {};
  \node at (0.8,-1.45) {(b)};
 \end{scope}
 \begin{scope}[shift={(8.0,0)}]
  \loopframe
  \draw[aux] (0,0) arc[start angle=180,end angle=0,radius=0.8];
  \draw[gluon] (0,0) arc[start angle=180,end angle=360,radius=0.8];
  \node[ctx] at (0.8,-0.8) {};
  \node[vtx] at (0,0) {}; \node[vtx] at (1.6,0) {};
  \node at (0.8,-1.45) {(c)};
 \end{scope}
\end{tikzpicture}
\caption{Two-loop first-order graphs with a quark-loop subdivergence. (a) Two background--auxiliary--gluon vertices joined by an auxiliary and a gluon propagator, with a quark loop on the gluon line. (b) The same with a magnetic vertex (square) and a mixed propagator. (c) The counterterm graph with the insertion of Fig.~\ref{fig:quarksector}(b) on the gluon line, which removes the subdivergence of (a).}
\label{fig:twoloop}
\end{figure}

Because $z$ is a pole, the renormalized auxiliary field depends on the renormalization scale through a mixing with the field strength. Holding the bare fields fixed in Eq.~\eqref{eq:Fbare} and using $\mu\,\dd g^2/\dd\mu=-2\epsilon g^2+O(g^4)$, we find $\mu\,\dd z^{(q)}/\dd\mu=-\frac23T_FN_f(-2\epsilon g^2)/(16\pi^2\eb)\to\frac43T_FN_f\,g^2/16\pi^2$, and hence
\begin{equation}
\mu\frac{\dd F}{\dd\mu}=\gamma_F\,f(A)+\dots,\qquad \gamma_F^{(q)}=-\frac{\alpha_sT_FN_f}{3\pi},
\label{eq:gammaF}
\end{equation}
where the dots denote the gluonic contribution, which is the same in theories with different numbers of flavours, and terms of higher order. Integrating at fixed coupling,
\begin{equation}
F(\mu')=F(\mu)-\frac{\alpha_sT_FN_f}{6\pi}\ln\frac{\mu'^2}{\mu^2}\,f(A).
\label{eq:Frun}
\end{equation}
The first-order form of the Lagrangian is preserved at every scale, but the field called $F$ at one scale is a mixture of $F$ and $f(A)$ at another. In the second-order theory there is no counterpart of Eqs.~\eqref{eq:gammaF} and \eqref{eq:Frun}, because there is no field $F$. The mixing \eqref{eq:zq} is also the counterterm for every quark-loop subdivergence on a gluon line in first-order graphs at two loops, as illustrated in Fig.~\ref{fig:twoloop}.

\section{Heavy-quark thresholds act on the auxiliary field}
\label{sec:threshold}

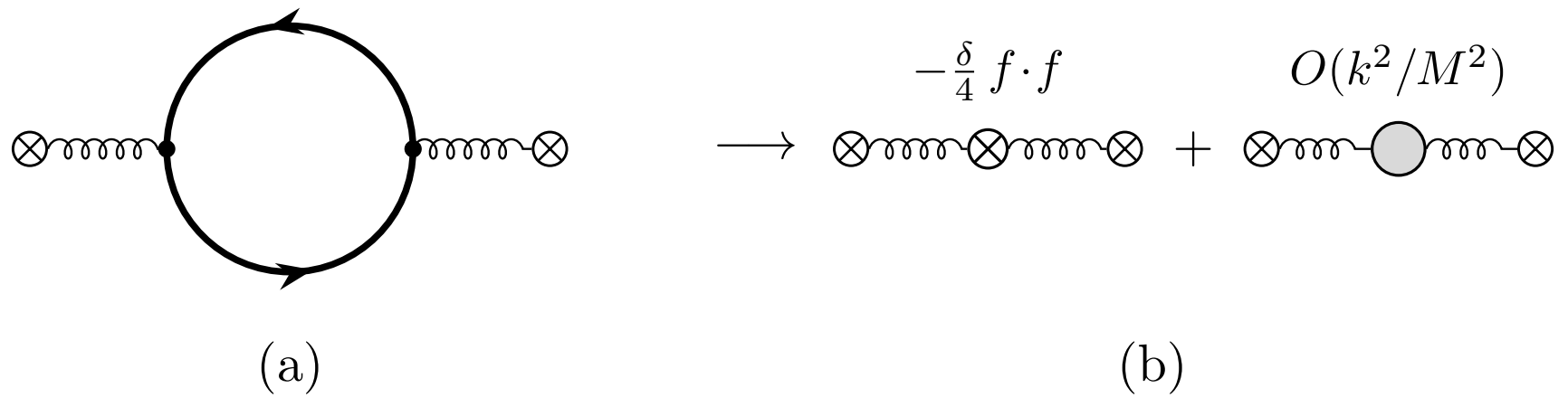
\begin{figure}[t]
\centering
\begin{tikzpicture}
  \begin{scope}
    \node[bg] (B1) at (-1.0,0) {}; \node[bg] (B2) at (2.8,0) {};
    \coordinate (v1) at (0,0); \coordinate (v2) at (1.8,0);
    \draw[gluon] (B1)--(v1); \draw[gluon] (v2)--(B2);
    \draw[heavy] (v2) arc[start angle=0,end angle=180,radius=0.9];
    \draw[heavy] (v1) arc[start angle=180,end angle=360,radius=0.9];
    \node[vtx] at (v1) {}; \node[vtx] at (v2) {};
    \node at (0.9,-1.6) {(a)};
  \end{scope}
  \begin{scope}[shift={(4.6,0)}]
    \node at (-0.3,0) {$\longrightarrow$};
    \node[bg] (C1) at (0.4,0) {}; \node[bg] (C2) at (2.4,0) {};
    \node[ctx] (X) at (1.4,0) {};
    \draw[gluon] (C1)--(X); \draw[gluon] (X)--(C2);
    \node[above] at (1.4,0.22) {\small $-\frac\delta4\,f\!\cdot\! f$};
    \node at (2.9,0) {$+$};
    \node[bg] (D1) at (3.4,0) {}; \node[bg] (D2) at (5.4,0) {};
    \node[blob,minimum size=11pt] (Y) at (4.4,0) {};
    \draw[gluon] (D1)--(Y); \draw[gluon] (Y)--(D2);
    \node[above] at (4.4,0.25) {\small $O(k^2/M^2)$};
    \node at (2.6,-1.6) {(b)};
  \end{scope}
\end{tikzpicture}
\caption{(a) The heavy-quark loop (thick line), which couples only to the gluon. (b) For $k^2\ll M^2$ it is replaced by the local term $-\frac\delta4f\cdot f$ (crossed circle) and by $1/M^2$ power corrections (blob).}
\label{fig:heavy}
\end{figure}

Consider $n_l$ light quarks and one heavy quark $h$ of mass $M$, and momenta $k^2\ll M^2$. At one loop in the heavy quark,
\begin{equation}
e^{i\Gamma_h[A]}=\int\mathcal Dh\,\mathcal D\bar h\;e^{i\int\bar h(i\slashed D-M)h}={\rm Det}\big(i\slashed D(A)-M\big),
\label{eq:det}
\end{equation}
a gauge-invariant functional of $A$ alone, independent of $F$. Its two-point part is Fig.~\ref{fig:heavy}(a), that is, Eq.~\eqref{eq:Piq} with $N_f\to1$ and $m\to M$. After $\MSbar$ renormalization in the full theory, expanding $\ln(M^2-x(1-x)k^2)=\ln M^2-x(1-x)k^2/M^2+O(k^4/M^4)$ and using $\int_0^1x(1-x)\,\dd x=\frac16$ and $\int_0^1x^2(1-x)^2\,\dd x=\frac1{30}$,
\begin{equation}
\Pi^R_h(k^2)=-\frac{\alpha_sT_F}{3\pi}\ln\frac{M^2}{\mu^2}+\frac{\alpha_sT_F}{15\pi}\,\frac{k^2}{M^2}+O\Big(\frac{k^4}{M^4}\Big).
\label{eq:PiRh}
\end{equation}
At order $M^0$ the heavy quark adds $-\frac14\Pi^R_h(0)\,f\cdot f$ to the light Lagrangian, which becomes
\begin{equation}
\begin{gathered}
\Lag_{\rm low}=\frac14F\cdot F-\frac12F\cdot f(A)-\frac\delta4f(A)\cdot f(A)+\sum_{\rm light}\bar\psi(i\slashed D-m)\psi+\dots,\\
\delta\equiv\Pi^R_h(0)=-\frac{\alpha_sT_F}{3\pi}\ln\frac{M^2}{\mu^2},
\end{gathered}
\label{eq:Llow}
\end{equation}
that is, $(\kappa,\lambda,\tau)=(1,1,\delta)$. The extra term is a second-order kinetic term. Expanded in fields it contains a cubic vertex with one derivative and a quartic vertex, the momentum-dependent interactions that the first-order formalism is designed to avoid. As in Section~\ref{sec:mixing}, no rescaling can remove it.

The extra term is removed, and the first-order form restored below the threshold, by the mixing \eqref{eq:mixmap}. Requiring $\tau'=\delta+2c-c^2=0$ and taking the root that vanishes with $\delta$,
\begin{equation}
c=1-\sqrt{1+\delta}=-\frac12\Pi^R_h(0)+O(\alpha_s^2)=\frac{\alpha_sT_F}{6\pi}\ln\frac{M^2}{\mu^2}+O(\alpha_s^2).
\label{eq:c}
\end{equation}
Explicitly, $c^2-2c-\delta=0$ has the roots $c=1\pm\sqrt{1+\delta}$, and only the lower sign vanishes as $\delta\to0$. By Eq.~\eqref{eq:mixmap} the mixing leaves $\kappa'=1$ and gives $\lambda'=1-c=\sqrt{1+\delta}$ and $\tau'=0$. The gluon rescaling with $s=(1+\delta)^{-1/2}$ then gives $(\kappa,s\lambda',s^2\tau')=(1,1,0)$, and the invariant changes from $\mathcal Z=1+\delta$ to $s^2\mathcal Z=1$. The light theory is again of first-order form,
\begin{equation}
\Lag_{\rm low}=\frac14F'\cdot F'-\frac12F'\cdot f_{g'}(A')+\sum_{\rm light}\bar\psi(i\slashed D'-m)\psi+\dots,
\label{eq:Lfinal}
\end{equation}
with
\begin{equation}
F=F'+c\,f_g(A),\qquad A'=\sqrt{1+\delta}\,A,\qquad g'=\frac{g}{\sqrt{1+\delta}},
\label{eq:map}
\end{equation}
and unchanged light-quark couplings, $gA=g'A'$. The map is unique once $\kappa'=1$ is imposed. In the background gauge, $Q=sQ'$ gives $\xi'=(1+\delta)\xi$, while $\D_\mu$ is unchanged because $gB=g'B'$.

Equation~\eqref{eq:map} gives $\alpha_s'=g'^2/4\pi=\alpha_s/(1+\delta)=\alpha_s(1-\delta+\dots)$, and with Eq.~\eqref{eq:Llow},
\begin{equation}
\alpha_s^{(n_l)}(\mu)=\alpha_s^{(n_l+1)}(\mu)\left[1+\frac{\alpha_sT_F}{3\pi}\ln\frac{M^2}{\mu^2}+O(\alpha_s^2)\right],
\label{eq:decoup}
\end{equation}
which is the standard one-loop decoupling relation \cite{Bernreuther:1982,Chetyrkin:1998}. The physical content of decoupling lies entirely in the invariant $\mathcal Z$, as it must. What the first-order formulation adds is the coefficient $c$, which records how the threshold is shared between the auxiliary field and the gluon field. It vanishes at $\mu=M$ at this order, so with matching at $\mu=M$ the first-order form is preserved without mixing at one loop, while for any other matching scale it is not.

The threshold mixing is tied to the ultraviolet mixing of Section~\ref{sec:mixing} in two ways. First, for one heavy flavour Eqs.~\eqref{eq:zq} and \eqref{eq:c} combine into
\begin{equation}
z_h+c=-\frac{\alpha_sT_F}{6\pi}\left[\frac1\eb-\ln\frac{M^2}{\mu^2}\right]=-\frac12\,\Pi_h(0),
\label{eq:zplusc}
\end{equation}
where $\Pi_h(0)$ is the unrenormalized heavy-quark loop at zero momentum. Its pole is the mixing that renormalizes $F$ in the full theory, and its logarithm is the mixing that survives below the threshold. Second, the scale dependence of $c$ is
\begin{equation}
\mu\frac{\dd c}{\dd\mu}=-\frac{\alpha_sT_F}{3\pi}+O(\alpha_s^2),
\label{eq:cRG}
\end{equation}
which is exactly the heavy flavour's share of $\gamma_F^{(q)}$ in Eq.~\eqref{eq:gammaF}. Using $F'=F-c\,f(A)$ from Eq.~\eqref{eq:map}, the light-theory auxiliary field obeys
\begin{equation}
\mu\frac{\dd F'}{\dd\mu}=\left(\gamma_F^{(n_l+1)}-\mu\frac{\dd c}{\dd\mu}\right)f(A)=-\frac{\alpha_sT_F\,n_l}{3\pi}\,f(A)=\gamma_F^{(n_l)}\,f(A),
\label{eq:Fprimerun}
\end{equation}
so the running of the auxiliary field decouples in the same way as $\beta_0$: below the threshold it runs with the light flavours only. Equivalently, by Eq.~\eqref{eq:Frun}, $c\,f(A)$ is the heavy flavour's part of the running of $F$ accumulated between $\mu$ and $M$, and $F'$ is the full-theory field with that part evaluated at $\mu=M$. Equations~\eqref{eq:zplusc}--\eqref{eq:Fprimerun} are the consistency conditions that link the two results, and they hold identically at one loop.

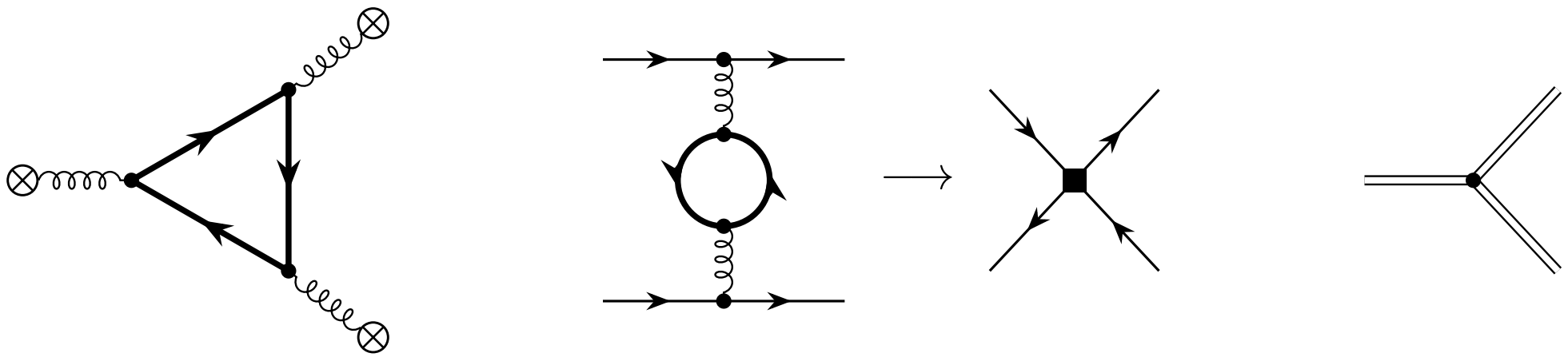
\begin{figure}[t]
\centering
\begin{tikzpicture}
 \begin{scope}
  \coordinate (T1) at (0,0); \coordinate (T2) at (1.3,0.75); \coordinate (T3) at (1.3,-0.75);
  \draw[heavy] (T1)--(T2); \draw[heavy] (T2)--(T3); \draw[heavy] (T3)--(T1);
  \node[bg] (B1) at (-0.9,0) {}; \node[bg] (B2) at (2.0,1.3) {}; \node[bg] (B3) at (2.0,-1.3) {};
  \draw[gluon] (B1)--(T1); \draw[gluon] (B2)--(T2); \draw[gluon] (B3)--(T3);
  \node[vtx] at (T1) {}; \node[vtx] at (T2) {}; \node[vtx] at (T3) {};
  \node at (0.6,-1.85) {(a)};
 \end{scope}
 \begin{scope}[shift={(3.9,0)}]
  \draw[quark] (0,1.0)--(1.0,1.0); \draw[quark] (1.0,1.0)--(2.0,1.0);
  \draw[quark] (0,-1.0)--(1.0,-1.0); \draw[quark] (1.0,-1.0)--(2.0,-1.0);
  \coordinate (u) at (1.0,0.38); \coordinate (d) at (1.0,-0.38);
  \draw[gluon] (1.0,1.0)--(u); \draw[gluon] (d)--(1.0,-1.0);
  \draw[heavy] (u) arc[start angle=90,end angle=270,radius=0.38];
  \draw[heavy] (d) arc[start angle=270,end angle=450,radius=0.38];
  \node[vtx] at (1.0,1.0) {}; \node[vtx] at (1.0,-1.0) {}; \node[vtx] at (u) {}; \node[vtx] at (d) {};
  \node at (2.6,0) {$\longrightarrow$};
  \coordinate (V) at (3.9,0);
  \draw[quark] (3.2,0.75)--(V); \draw[quark] (V)--(3.2,-0.75);
  \draw[quark] (4.6,-0.75)--(V); \draw[quark] (V)--(4.6,0.75);
  \node[mag,minimum size=5.5pt] at (V) {};
  \node at (2.6,-1.85) {(b)};
 \end{scope}
 \begin{scope}[shift={(10.2,0)}]
  \coordinate (V) at (0.9,0);
  \draw[aux] (0,0)--(V); \draw[aux] (V)--(1.6,0.75); \draw[aux] (V)--(1.6,-0.75);
  \node[vtx] at (V) {};
  \node at (0.9,-1.85) {(c)};
 \end{scope}
\end{tikzpicture}
\caption{Power corrections in the light first-order theory. (a) The heavy-quark triangle that fixes $C_2$. (b) Light quarks exchanging a gluon that carries the heavy loop; at order $k^2/M^2$ the exchange shrinks to the four-quark contact vertex (square) of Eq.~\eqref{eq:Llow6}. (c) The cubic, derivative-free vertex of the auxiliary field from $\mathcal O_2^F$.}
\label{fig:power}
\end{figure}

The power corrections come from the second term of Eq.~\eqref{eq:PiRh}, which gives, with $k^2\to-D^2$,
\begin{equation}
-\frac14\,\frac{\alpha_sT_F}{15\pi M^2}\,f\cdot(-D^2)f=\frac{a_h}{M^2}\,f\cdot D^2f,\qquad a_h\equiv\frac{\alpha_sT_F}{60\pi},
\label{eq:a}
\end{equation}
which for an Abelian theory with $\alpha_sT_F\to\alpha$ is the Uehling term \cite{Uehling:1935,Dunne:2005}. In the basis
\begin{equation}
\mathcal O_1=(D^\mu f_{\mu\nu})^a(D_\rho f^{\rho\nu})^a,\qquad \mathcal O_2=g\,f^{abc}f^a_\mu{}^\nu f^b_\nu{}^\rho f^c_\rho{}^\mu,
\label{eq:O12}
\end{equation}
the Bianchi identity $D^2f_{\alpha\beta}=D_\alpha D^\rho f_{\rho\beta}-D_\beta D^\rho f_{\rho\alpha}+O(gf^2)$ gives $f\cdot D^2f=-2\mathcal O_1$ up to a total derivative and a multiple of $\mathcal O_2$ (Appendix~\ref{app:stationarity}). The heavy quark therefore generates
\begin{equation}
\Lag_{\rm low}\supset\frac{C_1}{M^2}\mathcal O_1+\frac{C_2}{M^2}\mathcal O_2,\qquad C_1=-2a_h=-\frac{\alpha_sT_F}{30\pi},
\label{eq:C1}
\end{equation}
where $C_2$ follows from the heavy-quark triangle, Fig.~\ref{fig:power}(a) \cite{Novikov:1984,Cho:1995}; none of the statements below depends on its value. In second-order variables both operators contain derivatives and many gluon legs. In first-order variables we replace them by $\mathcal O_1^F=(D_\mu F^{\mu\nu})^a(D^\rho F_{\rho\nu})^a$ and $\mathcal O_2^F=gf^{abc}F^a_\mu{}^\nu F^b_\nu{}^\rho F^c_\rho{}^\mu$. The replacement changes the action only at order $1/M^4$, because $\mathcal O_i^F-\mathcal O_i$ vanishes at the leading-order solution $F=f(A)$ (Appendix~\ref{app:stationarity}). $\mathcal O_2^F$ is a cubic, derivative-free vertex of $F$. Before any use of the equations of motion, the light first-order theory at order $1/M^2$ is
\begin{equation}
\Lag_{\rm low}=\frac14F\cdot F-\frac12F\cdot f(A)+\sum_{\rm light}\bar\psi(i\slashed D-m)\psi+\frac{C_1}{M^2}\,\mathcal O_1^F+\frac{C_2}{M^2}\,\mathcal O_2^F+O\Big(\frac1{M^4}\Big).
\label{eq:Llow6b}
\end{equation}
For $\mathcal O_1^F$ the first-order equation of motion \eqref{eq:eomA}, now with the light quarks only, gives $\mathcal O_1^F=g^2J^a_\nu J^{a\nu}$ up to operators that vanish by the equations of motion and do not contribute to on-shell matrix elements \cite{Politzer:1980,Arzt:1995}. In the light theory Eq.~\eqref{eq:eomA} is no longer exact, because $\mathcal O_1^F$ depends on the gauge field through $D_\mu$, but these corrections are themselves of order $1/M^2$ and affect the reduction only at order $1/M^4$. The reduced form of the light first-order theory at order $1/M^2$ is therefore
\begin{equation}
\begin{split}
\Lag_{\rm low}={}&\frac14F\cdot F-\frac12F\cdot f(A)+\sum_{\rm light}\bar\psi(i\slashed D-m)\psi\\
&-\frac{\alpha_sT_F}{30\pi}\,\frac{g^2}{M^2}\,J^a_\nu J^{a\nu}+\frac{C_2}{M^2}\,g\,f^{abc}F^a_\mu{}^\nu F^b_\nu{}^\rho F^c_\rho{}^\mu+O\Big(\frac1{M^4}\Big),
\end{split}
\label{eq:Llow6}
\end{equation}
in the variables of Eq.~\eqref{eq:Lfinal} with primes dropped. The two forms \eqref{eq:Llow6b} and \eqref{eq:Llow6} are related by a redefinition of the gauge field of order $gJ/M^2$, given explicitly in Eq.~\eqref{eq:Alight} below, and have the same on-shell matrix elements. In the reduced form every interaction is momentum independent: the cubic $F\!AA$ vertex, the quark--gluon vertex, the four-quark contact term and the cubic $F$ vertex. Diagrammatically, the contact term is the gluon exchange of Fig.~\ref{fig:power}(b), in which the $k^2/M^2$ part of the heavy loop cancels the gluon pole. In the Abelian limit, $T_F\to1$, $g\to e$ and $J^\nu=\sum_iq_i\bar\psi_i\gamma^\nu\psi_i$, the contact term is $-(\alpha e^2/30\pi m^2)J_\nu J^\nu$. For static charges $J^0=\rho=q_1\delta^3(\mathbf x-\mathbf x_1)+q_2\delta^3(\mathbf x-\mathbf x_2)$, the Hamiltonian changes by $+(\alpha e^2/30\pi m^2)\int\rho^2$, whose cross term is $2q_1q_2\delta^3(\mathbf r)$. With $e^2=4\pi\alpha$ this gives the potential $\delta V=(4\alpha^2/15m^2)q_1q_2\,\delta^3(\mathbf r)$, which for opposite charges is the familiar contact part $-(4\alpha^2/15m^2)\delta^3(\mathbf r)$ of the Uehling potential \cite{Uehling:1935}.

We now ask what happens to the auxiliary field itself below the threshold. In the full theory the equation of motion of $F$ is algebraic, $F=f(A)$, and this is the content of the structural identities of the first-order formalism \cite{McKeon:2020}: Green's functions with an insertion of $F$ equal those with an insertion of the composite operator $f(A)$, up to contact terms. Because $F$ and $f(A)$ are off-shell quantities, the form the identity takes in the light theory depends on which of the two equivalent forms of the light Lagrangian is used. In the reduced form \eqref{eq:Llow6} the four-quark term does not involve $F$, so the field equation of $F$ is again algebraic and $F=f(A)+O(gf^2/M^2)$, the correction coming from $\mathcal O_2^F$. In the unreduced form \eqref{eq:Llow6b}, which is the form obtained directly from the heavy-quark loop, the power correction $\mathcal O_1^F$ contains derivatives of $F$, and the identity changes. In the rest of this section we work with Eq.~\eqref{eq:Llow6b}. With the convention that $X\cdot Y$ sums over all $\mu,\nu$, the variation of $\mathcal O_1^F$ is $\delta\mathcal O_1^F=2(D^\rho F_{\rho\nu})(D_\mu\delta F^{\mu\nu})\to-2(D_\mu D^\rho F_{\rho\nu})\,\delta F^{\mu\nu}$ after integration by parts, and only its antisymmetric part survives because $\delta F$ is antisymmetric. The equation of motion of $F$ is therefore
\begin{equation}
\frac12F_{\alpha\beta}-\frac12f_{\alpha\beta}-\frac{C_1}{M^2}\Big(D_\alpha D^\rho F_{\rho\beta}-D_\beta D^\rho F_{\rho\alpha}\Big)+(\mathcal O_2^F\ \text{term})=0 .
\label{eq:Feom6}
\end{equation}
Solving to order $1/M^2$, using the gluon equation \eqref{eq:eomA} for the light quarks, $D^\rho F_{\rho\beta}=-gJ_\beta$, whose $1/M^2$ corrections contribute here only at order $1/M^4$, and $-2C_1=\alpha_sT_F/15\pi$,
\begin{equation}
F_{\alpha\beta}=f_{\alpha\beta}(A)+\frac{g\,\alpha_sT_F}{15\pi M^2}\big(D_\alpha J_\beta-D_\beta J_\alpha\big)+O\Big(\frac{gf^2}{M^2},\frac1{M^4}\Big).
\label{eq:identity}
\end{equation}
The omitted term of order $gf^2/M^2$ comes from $\mathcal O_2^F$ and does not involve the light quarks. Below a heavy-quark threshold the auxiliary field is therefore the gluon field strength plus a contact term proportional to the covariant curl of the light-quark colour current, with a coefficient fixed by the heavy-quark loop. For Green's functions,
\begin{equation}
\Big\langle F_{\alpha\beta}(x)\,X\Big\rangle=\Big\langle f_{\alpha\beta}(x)\,X\Big\rangle+\frac{g\,\alpha_sT_F}{15\pi M^2}\Big\langle\big(D_\alpha J_\beta-D_\beta J_\alpha\big)(x)\,X\Big\rangle+\dots,
\label{eq:identityGF}
\end{equation}
for any product $X$ of light fields at other points, computed with Eq.~\eqref{eq:Llow6b}, so the structural identities of Ref.~\cite{McKeon:2020} acquire insertions of the curl of the light-quark current, suppressed by $1/M^2$. In pure Yang--Mills theory below the threshold ($n_l=0$) this term is absent and the deviation from $F=f(A)$ starts with $\mathcal O_2^F$.

We have verified Eq.~\eqref{eq:identity} in three independent ways. First, writing the power correction as $(a_h/M^2)F\cdot D^2F$ instead of $(C_1/M^2)\mathcal O_1^F$, which is allowed by the stationarity argument of Appendix~\ref{app:stationarity}, the equation of motion of $F$ gives $F=f-(4a_h/M^2)D^2F=f-(4a_h/M^2)D^2f+O(1/M^4)$. With the Bianchi identity, $D^2f_{\alpha\beta}=D_\alpha D^\rho f_{\rho\beta}-D_\beta D^\rho f_{\rho\alpha}+O(gf^2)$, and Eq.~\eqref{eq:eomA}, this is $-g(D_\alpha J_\beta-D_\beta J_\alpha)$, and since $4a_h=\alpha_sT_F/15\pi$ it reproduces Eq.~\eqref{eq:identity}.

Second, the covariant divergence of Eq.~\eqref{eq:identity} must reproduce the field equation of the light second-order theory. Applying $D^\alpha$, using $D^\alpha F_{\alpha\beta}=-gJ_\beta$ and the covariant conservation $D^\alpha J_\alpha=0$ of the light current, we find at this order
\begin{equation}
D^\alpha f_{\alpha\beta}=-g\,J_\beta-\frac{g\,\alpha_sT_F}{15\pi M^2}\,D^2J_\beta .
\label{eq:Uehlingeom}
\end{equation}
This is exactly the field equation that follows from the second-order light Lagrangian $-\frac14f\cdot f+(C_1/M^2)\mathcal O_1+gA\cdot J$, in which the variation of $(C_1/M^2)\mathcal O_1$ adds $(2C_1/M^2)D^2D^\alpha f_{\alpha\beta}=-(2C_1g/M^2)D^2J_\beta$ at this order. It is the non-Abelian form of the Uehling-corrected Maxwell equation.

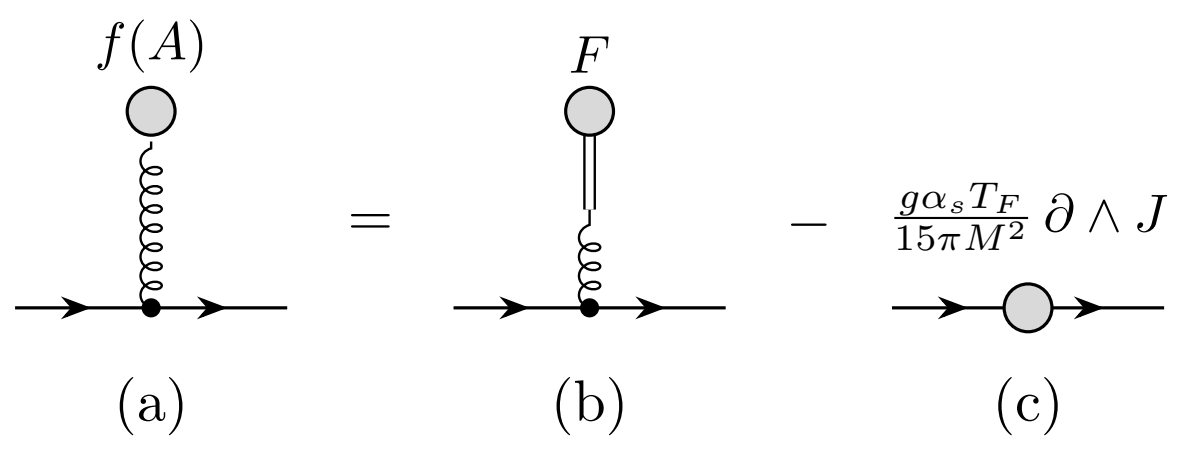
\begin{figure}[t]
\centering
\begin{tikzpicture}
 \newcommand{\qline}{\draw[quark] (0,0)--(0.9,0); \draw[quark] (0.9,0)--(1.8,0); \node[vtx] at (0.9,0) {};}
 \begin{scope}
  \qline
  \draw[gluon] (0.9,0)--(0.9,1.1);
  \node[blob] (O) at (0.9,1.3) {}; \node[above] at (0.9,1.42) {\small $f(A)$};
  \node at (0.9,-0.65) {(a)};
 \end{scope}
 \node at (2.35,0.55) {$=$};
 \begin{scope}[shift={(2.9,0)}]
  \qline
  \draw[gluon] (0.9,0)--(0.9,0.65); \draw[aux] (0.9,0.65)--(0.9,1.15);
  \node[blob] (O) at (0.9,1.3) {}; \node[above] at (0.9,1.42) {\small $F$};
  \node at (0.9,-0.65) {(b)};
 \end{scope}
 \node at (5.25,0.55) {$-$};
 \begin{scope}[shift={(5.8,0)}]
  \draw[quark] (0,0)--(0.9,0); \draw[quark] (0.9,0)--(1.8,0);
  \node[blob] at (0.9,0) {};
  \node[above] at (0.9,0.25) {\small $\frac{g\alpha_sT_F}{15\pi M^2}\,\partial\wedge J$};
  \node at (0.9,-0.65) {(c)};
 \end{scope}
\end{tikzpicture}
\caption{The identity \eqref{eq:identity} for a light-quark matrix element below a heavy-quark threshold. (a) The field strength $f(A)$ (shaded circle) is reached through a gluon line and contains the heavy-quark contact term, as in the full theory, Eq.~\eqref{eq:Afull}. (b) The auxiliary field is reached through the mixed propagator and is a pure gluon pole, Eq.~\eqref{eq:Alight}. (c) The difference is a contact insertion of the curl of the light-quark current, with no propagator.}
\label{fig:identity}
\end{figure}

Third, Eq.~\eqref{eq:identity} can be checked against the full theory, where $F=f(A)$ holds exactly. Consider a classical light-quark source $J$ at linear order in $g$, with $\partial\cdot J=0$. In the full theory the heavy loop modifies the gluon kinetic term to $-\frac14f\cdot[1+\Pi^R_h(-\partial^2)]f$, so $[1+\Pi^R_h(-\partial^2)]\Box A=-gJ$ in a transverse gauge. With Eq.~\eqref{eq:PiRh}, $\Pi^R_h(-\partial^2)=\delta-(\alpha_sT_F/15\pi M^2)\Box$, and
\begin{equation}
A_{\rm full}=-g(1-\delta)\,\Box^{-1}J-\frac{g\,\alpha_sT_F}{15\pi M^2}\,J+O\Big(\frac1{M^4}\Big).
\label{eq:Afull}
\end{equation}
The second term is a contact term: the heavy-quark loop cancels the gluon pole. In the light theory, Eq.~\eqref{eq:identity} at linear order says $F=\partial\wedge\tilde A$ with $\tilde A=A+(g\alpha_sT_F/15\pi M^2)J$, and $\partial_\mu F^{\mu\nu}=-gJ^\nu$ then gives $\Box\tilde A=-gJ$. Hence
\begin{equation}
F_{\rm light}=-g\,\Box^{-1}(\partial\wedge J),\qquad A_{\rm light}=-g\,\Box^{-1}J-\frac{g\,\alpha_sT_F}{15\pi M^2}\,J .
\label{eq:Alight}
\end{equation}
The contact term of $A_{\rm light}$ coincides with that of $A_{\rm full}$; the factor $1-\delta$ is the rescaling of Eq.~\eqref{eq:map}. This is a non-trivial check, since the light theory knows nothing about the heavy loop except through $C_1$. It also gives Eq.~\eqref{eq:identity} a direct physical meaning, shown in Fig.~\ref{fig:identity}. The field strength $f(A)$ of the light theory carries the short-distance effect of the heavy quark in its matrix elements, as it must, since it reproduces the full theory. The auxiliary field $F$ does not: its single-gluon matrix element between light quarks is a pure gluon pole. This statement does not depend on the form of the light Lagrangian. In the reduced form \eqref{eq:Llow6} the field redefinition $A\to\tilde A$ removes the contact term from the gauge field, $F=f(\tilde A)$ is the same pure pole, and the heavy quark's short-distance effect sits explicitly in the four-quark contact vertex. Below a threshold the first-order formulation thus separates the long-distance exchange, carried by $F$, from the short-distance heavy-quark effect, carried by the contact term, and Eq.~\eqref{eq:identity} is the operator statement of this separation. The coefficient is $g\alpha_sT_F/15\pi M^2$ with $\alpha_sT_F/15\pi=3.4\times10^{-3}$ at the charm threshold and $2.2\times10^{-3}$ at the bottom threshold, for the inputs of Table~\ref{tab:summary}.

\begin{table}[t]
\centering
\small
\renewcommand{\arraystretch}{1.45}
\begin{tabular}{lll}
\toprule
Quantity & Result & Equation\\
\midrule
quark mixing of $F$ with $f(A)$ & $z^{(q)}=-\alpha_sT_FN_f/(6\pi\eb)$ & \eqref{eq:zq}\\
rescalings & $\delta Z_A^{(q)}=\delta Z_B^{(q)}=2z^{(q)}$, \ $\delta Z_F^{(q)}=0$ & \eqref{eq:zq}\\
running of $F$ (quark part) & $\mu\,\dd F/\dd\mu=-(\alpha_sT_FN_f/3\pi)\,f(A)$ & \eqref{eq:gammaF}\\
threshold mixing & $c=(\alpha_sT_F/6\pi)\ln(M^2/\mu^2)$ & \eqref{eq:c}\\
sum rule & $z_h+c=-\frac12\Pi_h(0)$ & \eqref{eq:zplusc}\\
running of $c$ & $\mu\,\dd c/\dd\mu=-\alpha_sT_F/3\pi$ & \eqref{eq:cRG}\\
four-quark contact term & $-(\alpha_sT_F/30\pi)\,g^2J\cdot J/M^2$ & \eqref{eq:Llow6}\\
field-strength identity, form \eqref{eq:Llow6b} & $F=f+(g\alpha_sT_F/15\pi M^2)(D\wedge J)$ & \eqref{eq:identity}\\
\bottomrule
\end{tabular}
\caption{Summary of the first-order results for the quark sector and for heavy-quark thresholds at one loop. For the charm and bottom thresholds at $\mu=M$, with $m_c=1.273$~GeV and $m_b=4.183$~GeV \cite{PDG:2024}, and with $\alpha_s^{(4)}(m_c)=0.319$ and $\alpha_s^{(5)}(m_b)=0.212$ obtained by one-loop running and one-loop matching from $\alpha_s^{(5)}(M_Z)=0.1180$ \cite{PDG:2024} (higher-order running gives larger values, especially at $m_c$, so these numbers are indicative), the four-quark coupling $2\alpha_s^2T_F/15M^2$ is $4.2\times10^{-3}$ and $1.7\times10^{-4}$~GeV$^{-2}$.}
\label{tab:summary}
\end{table}

\section{Conclusions}
\label{sec:concl}

Quarks couple only to the gauge field, and this simple fact changes the first-order formulation of QCD in three ways that have no counterpart in the second-order theory. First, the quark divergence is a multiple of $f(A)\cdot f(A)$, which the first-order Lagrangian does not contain and no rescaling can produce. The quark sector is renormalized entirely by the mixing $F_0=F+z\,f(A)$ known from pure Yang--Mills theory \cite{Frenkel:2017}, with quark coefficient $z^{(q)}=-\alpha_sT_FN_f/(6\pi\eb)$, a rescaling of the gluon field equal to the quark part of $Z_B$, and no rescaling of $F$. The mixing is independent of the gauge parameter and of the quark masses, preserves the background value $f(B)$ of the auxiliary field, and makes the renormalized auxiliary field run into the gluon field strength at the rate $\alpha_sT_FN_f/3\pi$. Second, a heavy-quark threshold acts on the auxiliary field through the same mechanism. The light theory is brought back to first-order form by a finite mixing $c=(\alpha_sT_F/6\pi)\ln(M^2/\mu^2)$ and a gluon rescaling that encodes the decoupling of $\alpha_s$. The two mixings combine into $-\frac12$ times the zero-momentum heavy-quark loop, the scale dependence of $c$ is exactly the heavy flavour's share of the running of $F$, so that below the threshold $F$ runs with the light flavours only, and at order $1/M^2$ the light theory keeps momentum-independent vertices only. Third, below the threshold, with the leading power correction kept as the gauge-sector operator $\mathcal O_1^F$, the auxiliary field is no longer the field strength. It satisfies $F=f(A)+(g\alpha_sT_F/15\pi M^2)(D_\alpha J_\beta-D_\beta J_\alpha)$, so the structural identities of the first-order formalism acquire insertions of the curl of the light-quark current. Matching to the full theory shows that $f(A)$ carries the heavy-quark contact term while $F$ is a pure gluon pole: the first-order variables separate the long-distance exchange from the short-distance effect of the heavy quark.

These results concern the auxiliary field itself and are not fixed by the equivalence of the first-order and second-order theories, which constrains only the invariant combination $\mathcal Z$ of the gauge couplings. They have two direct uses. The mixing $z^{(q)}$ is the counterterm of every quark-loop subdivergence in first-order graphs, Fig.~\ref{fig:twoloop}, and is therefore needed for the quark contributions to $\beta_1$ in this formalism. The threshold mixing $c$ is needed whenever a first-order calculation is matched across a heavy-quark threshold. The limits are also clear. Everything here is one loop in the quarks. At two loops the hard region contains gluon lines, through the mixed propagators the auxiliary field enters the matching directly, and the two-loop values of $z$ and $c$ and the corresponding correction to Eq.~\eqref{eq:identity} require a genuine first-order two-loop calculation, which we leave for future work. The reduced Lagrangian \eqref{eq:Llow6} was obtained with the equations of motion and describes on-shell matrix elements; the identity \eqref{eq:identity} refers to the unreduced form \eqref{eq:Llow6b}, and after the reduction the contact effect moves from the gauge field into the four-quark vertex while $F=f(A)$ up to the $\mathcal O_2^F$ term.

\acknowledgments

This work is an extension of the author's earlier study, ``Renormalization of Massless Quarks in the First-Order Background Field Formalism of QCD'', carried out under the supervision of Prof.~M.~Arshad Momen and presented as an oral talk at the International Conference on Physics, University of Dhaka, Bangladesh, in April 2026. The present paper takes the next step: it asks what quark loops do to the auxiliary field itself, extends the analysis to massive quarks and heavy-quark thresholds, and follows the auxiliary field below a threshold.

The idea for both studies grew out of the author's M.Sc.\ thesis. The author is deeply grateful to Prof.~M.~Arshad Momen (Department of Physical Sciences, Independent University, Bangladesh), co-supervisor of that thesis and the author's course teacher in quantum field theory, from whom the author learned the methods used here; his guidance, encouragement and many helpful discussions on the first-order formalism shaped this work from the original idea to its present form. The author thanks Dr.~Md Tanjib Atique Khan, supervisor of the M.Sc. thesis, for his support during the thesis work, and the Department of Theoretical Physics, University of Dhaka, for providing the academic environment in which this research was carried out. The author acknowledges the use of Claude Opus 5.5 (Anthropic) for cross-checks of the calculations, suggestions on figure presentation, reference searches and grammatical corrections. The author has checked all the results and takes full responsibility for the material presented here.

\appendix

\section{Conventions and the extraction of poles}
\label{app:conv}

We use $\{\gamma^\mu,\gamma^\nu\}=2\eta^{\mu\nu}$ and, in $d$ dimensions, $\gamma^\mu\gamma_\mu=d$ and $\gamma^\mu\gamma^\alpha\gamma_\mu=(2-d)\gamma^\alpha$, with $\tr\mathbf 1=4$. Denominators are combined with $1/(AB)=\int_0^1\dd x\,[xA+(1-x)B]^{-2}$, the basic integral is Eq.~\eqref{eq:I2}, and $\int\ell^2(\ell^2-\Delta)^{-2}=\frac d{d-2}\Delta\int(\ell^2-\Delta)^{-2}$. Ultraviolet poles of the gluonic and vertex graphs are extracted by large-momentum projection: if $\mathcal I_{-4}(\ell)$ is the part of the integrand homogeneous of degree $-4$ in the loop momentum, then
\begin{equation}
\mu^{2\epsilon}\!\int\!\frac{\dd^d\ell}{(2\pi)^d}\,\mathcal I(\ell)\UV=\frac{i}{16\pi^2\eb}\,\Big\langle(\ell^2)^2\,\mathcal I_{-4}(\ell)\Big\rangle,
\label{eq:UVproj}
\end{equation}
where $\langle\cdots\rangle$ is the angular average, $\ell_\mu\ell_\nu\to\frac14\ell^2\eta_{\mu\nu}$ and $\ell_\mu\ell_\nu\ell_\rho\ell_\sigma\to\frac1{24}(\ell^2)^2(\eta_{\mu\nu}\eta_{\rho\sigma}+\eta_{\mu\rho}\eta_{\nu\sigma}+\eta_{\mu\sigma}\eta_{\nu\rho})$. Terms of other degrees and scaleless integrals give no pole.

\section{Feynman rules and counterterms}
\label{app:rules}

Figure~\ref{fig:rules} collects the propagators and the vertices of the first-order background-field theory that are used in this paper, and Fig.~\ref{fig:ct} the counterterm insertions. The analytic expressions are given in Eqs.~\eqref{eq:DQQ}--\eqref{eq:DGG} and \eqref{eq:Vertices}, together with $\vev{c^a\bar c^b}=i\delta^{ab}/p^2$ and $\vev{\psi\bar\psi}=i(\slashed p+m)/(p^2-m^2)$. Vertices with two background legs arise from the non-Abelian part $gB\times B$ of $f(B)$ in the magnetic term, from the term quadratic in $B$ in the gauge-fixing Lagrangian, and from the ghost Lagrangian. At one loop they appear only in tadpole graphs with a single massless propagator, which are scaleless and vanish in dimensional regularization, so their explicit form is not needed here; they are shown in Fig.~\ref{fig:vanishing}. In the first-order theory there is no $BBQQ$ vertex from the kinetic term. Its second-order counterpart is generated by the local part of the auxiliary propagator, Fig.~\ref{fig:dictionary}.

\begin{figure}[!htbp]
\centering
\begin{tikzpicture}
 \begin{scope}
  \coordinate (V) at (0.8,0);
  \node[bg] (BL) at (0,0) {}; \node[bg] (BR) at (1.6,0) {};
  \draw[gluon] (BL)--(V); \draw[gluon] (V)--(BR);
  \draw[gluon] (V) arc[start angle=270,end angle=-90,radius=0.5];
  \node[mag] at (V) {};
  \node at (0.8,-1.25) {(a)};
 \end{scope}
 \begin{scope}[shift={(3.0,0)}]
  \coordinate (V) at (0.8,0);
  \node[bg] (BL) at (0,0) {}; \node[bg] (BR) at (1.6,0) {};
  \draw[gluon] (BL)--(V); \draw[gluon] (V)--(BR);
  \draw[gluon] (V) arc[start angle=270,end angle=-90,radius=0.5];
  \node[gfv] at (V) {};
  \node at (0.8,-1.25) {(b)};
 \end{scope}
 \begin{scope}[shift={(6.0,0)}]
  \coordinate (V) at (0.8,0);
  \node[bg] (BL) at (0,0) {}; \node[bg] (BR) at (1.6,0) {};
  \draw[gluon] (BL)--(V); \draw[gluon] (V)--(BR);
  \draw[ghost] (V) arc[start angle=270,end angle=-90,radius=0.5];
  \node[vtx] at (V) {};
  \node at (0.8,-1.25) {(c)};
 \end{scope}
 \begin{scope}[shift={(9.0,0)}]
  \coordinate (V1) at (0.4,0); \coordinate (V2) at (1.6,0);
  \node[bg] (BL) at (-0.3,0.55) {}; \node[bg] (BR) at (2.3,0.55) {};
  \draw[gluon] (BL)--(V1); \draw[gluon] (V2)--(BR);
  \draw[aux] (V1)--(V2); \node[above] at (1.0,0.02) {\scriptsize local};
  \draw[gluon] (V1) .. controls (0.4,-1.0) and (1.6,-1.0) .. (V2);
  \node[vtx] at (V1) {}; \node[vtx] at (V2) {};
  \node at (1.0,-1.25) {(d)};
 \end{scope}
\end{tikzpicture}
\caption{One-loop graphs that vanish in dimensional regularization because they are scaleless: tadpoles from (a) the $BB$ part of the magnetic coupling, (b) the $BB$ part of the gauge-fixing term, (c) the ghost Lagrangian, and (d) the local part $2iI$ of the auxiliary propagator in topology D1, which closes a single massless gluon line.}
\label{fig:vanishing}
\end{figure}
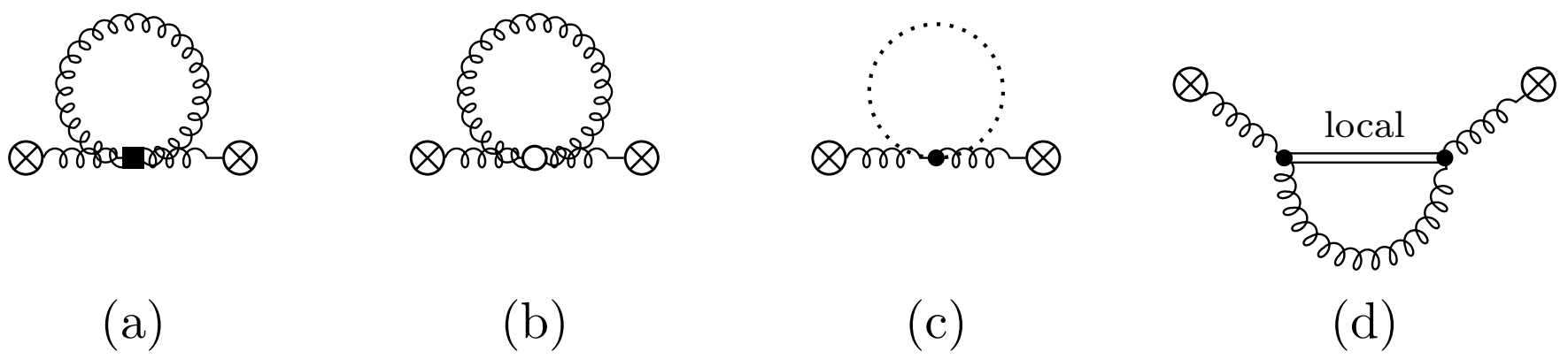

\begin{figure}[htbp]
\centering
\resizebox{\textwidth}{!}{\begin{tikzpicture}[every node/.style={font=\small}]
 \def\yA{0} \def\yB{-1.0} \def\yC{-2.0}
 \draw[gluon] (0,\yA)--(1.8,\yA); \node[right] at (2.1,\yA) {$\vev{Q^a_\mu Q^b_\nu}=\delta^{ab}D_{\mu\nu}(p)$, Eq.~\eqref{eq:DQQ}};
 \draw[aux] (8,\yA)--(9.8,\yA); \node[right] at (10.1,\yA) {$\vev{G^a_{\mu\nu}G^b_{\rho\sigma}}=2i\delta^{ab}(I-\Pi^L)$};
 \draw[aux] (0,\yB)--(0.9,\yB); \draw[gluon] (0.9,\yB)--(1.8,\yB); \node[right] at (2.1,\yB) {$\vev{G^a_{\mu\nu}Q^b_\rho}=-\delta^{ab}\dfrac{p_\mu\eta_{\nu\rho}-p_\nu\eta_{\mu\rho}}{p^2}$};
 \draw[quark] (8,\yB)--(9.8,\yB); \node[right] at (10.1,\yB) {$\dfrac{i(\slashed p+m)}{p^2-m^2}$};
 \draw[ghost] (0,\yC)--(1.8,\yC); \node[right] at (2.1,\yC) {$\vev{c^a\bar c^b}=\dfrac{i\delta^{ab}}{p^2}$};
 \newcommand{\vblock}[4]{
   \begin{scope}[shift={(#1,#2)}] #3 \end{scope}
   \node[right] at ($(#1,#2)+(2.1,0)$) {#4};}
 \vblock{0.4}{-3.6}{\coordinate (v) at (0.5,-0.15); \draw[quark] (0,-0.5)--(v); \draw[quark] (v)--(1.0,-0.5); \draw[gluon] (v)--(0.5,0.55); \node[vtx] at (v) {};}{$ig\gamma_\mu T^a$ (quantum $Q$)}
 \vblock{8.4}{-3.6}{\coordinate (v) at (0.5,-0.15); \draw[quark] (0,-0.5)--(v); \draw[quark] (v)--(1.0,-0.5); \node[bg] (b) at (0.5,0.6) {}; \draw[gluon] (b)--(v); \node[vtx] at (v) {};}{$ig\gamma_\mu T^a$ (background $B$)}
 \vblock{0.4}{-5.4}{\coordinate (v) at (0.5,-0.1); \node[bg] (b) at (0.5,0.6) {}; \draw[gluon] (b)--(v); \draw[aux] (v)--(0,-0.6); \draw[gluon] (v)--(1.0,-0.6); \node[vtx] at (v) {};}{GBQ: $-igf^{abc}\eta_{\mu\lambda}\eta_{\nu\rho}|_{[\mu\nu]}$}
 \vblock{8.4}{-5.4}{\coordinate (v) at (0.5,-0.1); \node[bg] (b) at (0.5,0.6) {}; \draw[gluon] (b)--(v); \draw[gluon] (v)--(0,-0.6); \draw[gluon] (v)--(1.0,-0.6); \node[mag] at (v) {};}{magnetic: Eq.~\eqref{eq:Vertices}}
 \vblock{0.4}{-7.2}{\coordinate (v) at (0.5,-0.1); \node[bg] (b) at (0.5,0.6) {}; \draw[gluon] (b)--(v); \draw[gluon] (v)--(0,-0.6); \draw[gluon] (v)--(1.0,-0.6); \node[gfv] at (v) {};}{gauge fixing: Eq.~\eqref{eq:Vertices}}
 \vblock{8.4}{-7.2}{\coordinate (v) at (0.5,-0.1); \node[bg] (b) at (0.5,0.6) {}; \draw[gluon] (b)--(v); \draw[ghost] (v)--(0,-0.6); \draw[ghost] (v)--(1.0,-0.6); \node[vtx] at (v) {};}{ghost: $[B^a_\lambda\bar c^bc^c]=gf^{abc}(p+p')_\lambda$}
 \vblock{0.4}{-9.0}{\coordinate (v) at (0.5,-0.1); \draw[aux] (0.5,0.6)--(v); \draw[gluon] (v)--(0,-0.6); \draw[gluon] (v)--(1.0,-0.6); \node[vtx] at (v) {};}{$GQQ$: $-igf^{abc}\eta_{\mu\rho}\eta_{\nu\sigma}|_{[\mu\nu]}$ (two loops)}
\end{tikzpicture}}
\caption{Propagators and vertices of first-order background-field QCD. Curly lines: quantum gluon $Q$; double lines: auxiliary field $G$; half-double lines: mixed propagator; dotted lines: ghosts; crossed circles: background field $B$. Momentum $p$ flows from the right end to the left end of each propagator.}
\label{fig:rules}

\vspace{1.2em}
\begin{tikzpicture}
  \begin{scope}
    \node[ctx] (X) at (1,0) {};
    \draw[quark] (0,0)--(X); \draw[quark] (X)--(2,0);
    \node at (1,-0.7) {\small $i\big[\delta Z_\psi\slashed p-(\delta Z_\psi+\delta Z_m)m\big]$};
  \end{scope}
  \begin{scope}[shift={(4.6,0)}]
    \node[ctx] (X) at (1,0) {};
    \draw[quark] (0,0)--(X); \draw[quark] (X)--(2,0);
    \node[bg] (b) at (1,1.0) {}; \draw[gluon] (b)--(X);
    \node at (1,-0.7) {\small $ig\,\delta Z_{1B}\gamma_\mu T^a$};
  \end{scope}
  \begin{scope}[shift={(9.0,0)}]
    \node[bg] (b1) at (0,0) {}; \node[ctx] (X) at (1.2,0) {}; \node[bg] (b2) at (2.4,0) {};
    \draw[gluon] (b1)--(X); \draw[gluon] (X)--(b2);
    \node at (1.2,-0.7) {\small $i\,\delta Z_B\,\delta^{ab}(k_\mu k_\nu-k^2\eta_{\mu\nu})$};
  \end{scope}
\end{tikzpicture}
\caption{Counterterm insertions from Eq.~\eqref{eq:dL} used in Appendices~\ref{sec:quark} and \ref{sec:twopoint}: quark two-point function, background quark--gluon vertex, and background two-point function.}
\label{fig:ct}
\end{figure}

\section{Quark self-energy and the background quark--gluon vertex}
\label{sec:quark}

These graphs verify $Z_{1B}=Z_\psi$, that is Eq.~\eqref{eq:ZgZB} in the quark sector, which is used in Eq.~\eqref{eq:Fbare}.

\paragraph{Self-energy.} The quark couples only to $Q$, and $\vev{QQ}$ is the second-order propagator, so the self-energy is the single graph of Fig.~\ref{fig:selfenergy}. With $S^{-1}=\slashed p-m-\Sigma$,
\begin{equation}
\Sigma(p)=-ig^2C_F\mu^{2\epsilon}\!\int\!\frac{\dd^dk}{(2\pi)^d}\,\frac{\gamma^\mu(\slashed p-\slashed k+m)\gamma^\nu}{k^2\left[(p-k)^2-m^2\right]}\left[\eta_{\mu\nu}-(1-\xi)\frac{k_\mu k_\nu}{k^2}\right].
\label{eq:Sigma1}
\end{equation}
In the Feynman-gauge part, $\gamma^\mu(\slashed p-\slashed k+m)\gamma_\mu=(2-d)(\slashed p-\slashed k)+dm$; after Feynman parametrization with $\ell=k-(1-x)p$ the pole is $C_F\K(-\slashed p+4m)$. In the longitudinal part we write $\slashed k=(\slashed p-m)-(\slashed p-\slashed k-m)$; the only divergent term is $-(\slashed p-m)\int(k^2)^{-2}$, with pole $(1-\xi)C_F\K(\slashed p-m)$. Hence
\begin{equation}
\Sigma\UV=C_F\K\left[-\xi\slashed p+(3+\xi)m\right],\qquad Z_\psi=1-\xi\,\frac{g^2C_F}{16\pi^2\eb},\qquad Z_m=1-\frac{3g^2C_F}{16\pi^2\eb}.
\label{eq:Sigmapole}
\end{equation}

\paragraph{Vertex.} Background gauge invariance gives the Abelian-like Ward identity $q^\mu\Lambda_\mu(p',p)=-\Sigma(p')+\Sigma(p)$, which with Eq.~\eqref{eq:Sigmapole} predicts $\Lambda_\mu\UV=\xi C_F\K\gamma_\mu$. The graphs are shown in Fig.~\ref{fig:vertex}. In graph (a) the colour factor is $T^bT^aT^b=(C_F-\frac12C_A)T^a$, and the leading numerator is $\gamma^\rho\slashed k\gamma_\mu\slashed k\gamma^\sigma P_{\rho\sigma}\to\xi k^2\gamma_\mu$, so $\Lambda^{(a)}\UV=\xi(C_F-\frac12C_A)\K\gamma_\mu$. The missing $\frac12\xi C_A\K\gamma_\mu$ must come from the graphs in which the background couples to the gluons. In the first-order formalism these are (b1), (b1$'$), (b2) and (b3); by the dictionary of Fig.~\ref{fig:dictionary} their background vertices add up to Eq.~\eqref{eq:Abbottvertex}, with colour factor $f^{abc}T^cT^b=-\frac i2C_AT^a$. The magnetic vertex is proportional to the background momentum, so (b2) is ultraviolet finite. For the others, contracting the leading parts of the vertices with $P_{\rho\sigma}=\eta_{\rho\sigma}-(1-\xi)\ell_\rho\ell_\sigma/\ell^2$ and averaging over directions gives
\begin{equation}
\Lambda^{(b1)+(b1')}\UV=\frac34(1+\xi)C_A\K\gamma_\mu,\qquad \Lambda^{(b2)}\UV=0,\qquad \Lambda^{(b3)}\UV=-\frac14(3+\xi)C_A\K\gamma_\mu .
\label{eq:Lambda}
\end{equation}
The auxiliary-field graphs reproduce the conventional non-Abelian vertex correction \cite{Pascual:1984,Davydychev:2001}, and the gauge-fixing graph, which has no conventional analogue, converts it into the background-field result: the sum is $\frac12\xi C_A\K\gamma_\mu$, Fig.~\ref{fig:vertexsum}. Adding graph (a), $\Lambda_\mu\UV=\xi C_F\K\gamma_\mu$, so
\begin{equation}
Z_{1B}=1-\xi\,\frac{g^2C_F}{16\pi^2\eb}=Z_\psi
\label{eq:Z1B}
\end{equation}
for every $\xi$, as required by Eq.~\eqref{eq:ZgZB}.

\begin{figure}[t]
\centering
\begin{tikzpicture}
  \coordinate (a) at (0,0); \coordinate (v1) at (1.2,0); \coordinate (v2) at (3.2,0); \coordinate (b) at (4.4,0);
  \draw[quark] (a)--(v1); \draw[quark] (v1)--(v2); \draw[quark] (v2)--(b);
  \draw[gluon] (v1) arc[start angle=180,end angle=0,radius=1.0];
  \node[vtx] at (v1) {}; \node[vtx] at (v2) {};
  \draw[mom] (1.75,1.25) arc[start angle=120,end angle=60,radius=0.9];
  \node at (2.2,1.55) {$k$};
  \draw[mom] (0.25,-0.32)--(0.95,-0.32); \node at (0.6,-0.6) {$p$};
  \draw[mom] (1.85,-0.32)--(2.55,-0.32); \node at (2.2,-0.6) {$p-k$};
\end{tikzpicture}
\caption{One-loop quark self-energy. The curly line is the quantum gluon propagator $\vev{QQ}$ of Eq.~\eqref{eq:DQQ}, which in the first-order theory coincides with the second-order propagator.}
\label{fig:selfenergy}
\end{figure}
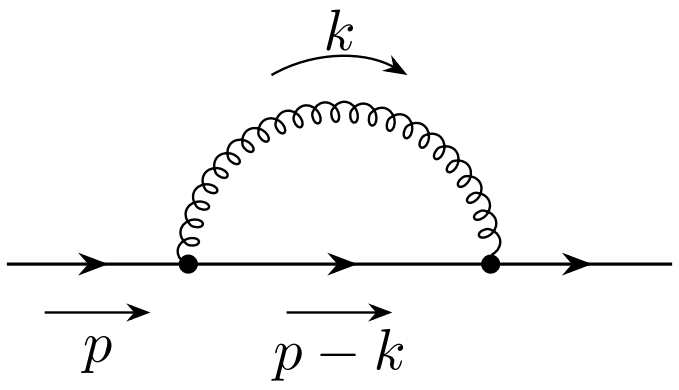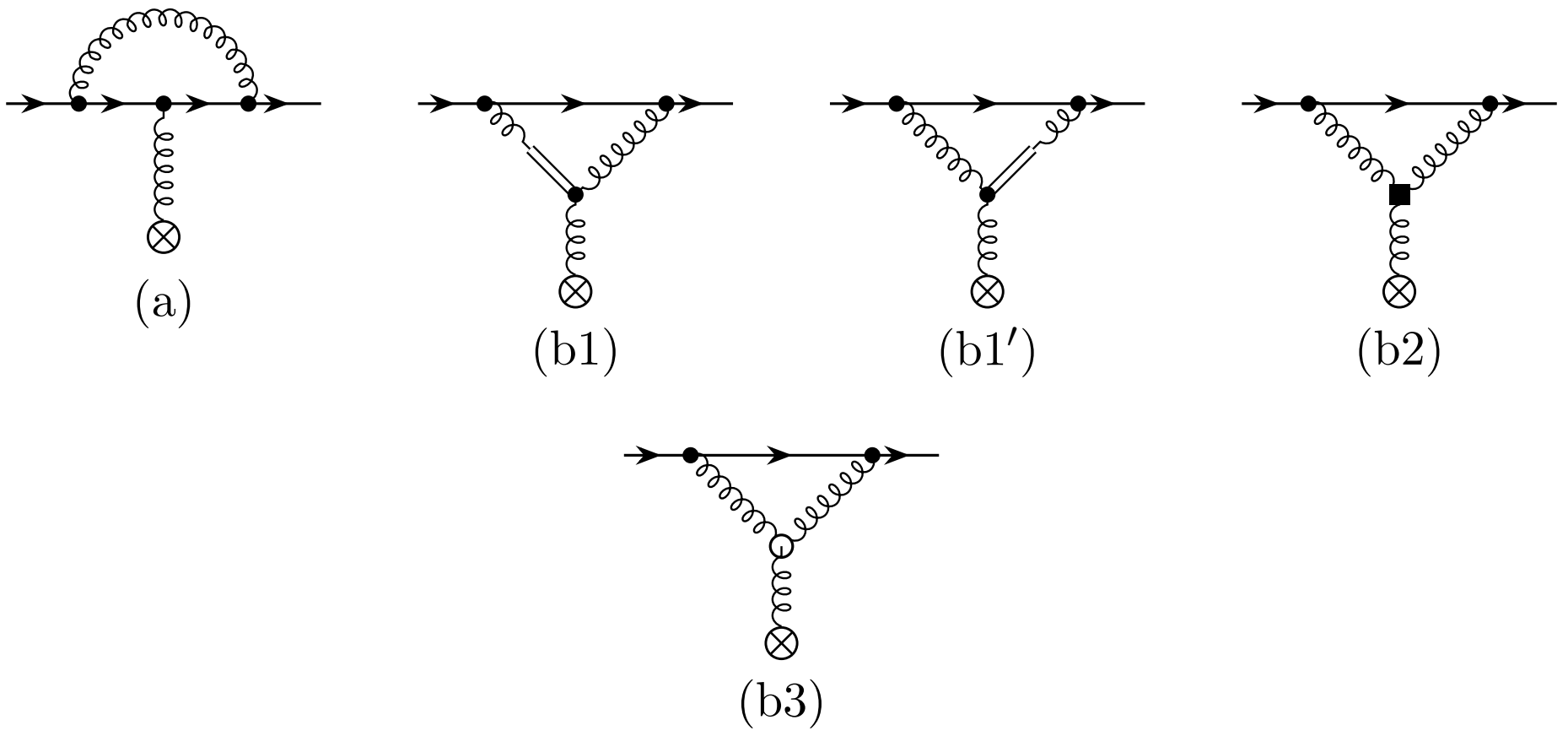
\begin{figure}[t]
\centering
\begin{tikzpicture}[scale=0.95]
  \begin{scope}
    \draw[quark] (0,0)--(0.6,0); \draw[quark] (0.6,0)--(1.3,0); \draw[quark] (1.3,0)--(2.0,0); \draw[quark] (2.0,0)--(2.6,0);
    \draw[gluon] (0.6,0) arc[start angle=180,end angle=0,radius=0.7];
    \node[vtx] at (0.6,0) {}; \node[vtx] at (2.0,0) {}; \node[vtx] at (1.3,0) {};
    \node[bg] (B) at (1.3,-1.1) {}; \draw[gluon] (B)--(1.3,0);
    \node at (1.3,-1.65) {(a)};
  \end{scope}
  \begin{scope}[shift={(3.4,0)}]
    \draw[quark] (0,0)--(0.55,0); \draw[quark] (0.55,0)--(2.05,0); \draw[quark] (2.05,0)--(2.6,0);
    \coordinate (V) at (1.3,-0.75);
    \draw[gluon] (0.55,0)--($(0.55,0)!0.5!(V)$); \draw[aux] ($(0.55,0)!0.5!(V)$)--(V);
    \draw[gluon] (2.05,0)--(V);
    \node[vtx] at (0.55,0) {}; \node[vtx] at (2.05,0) {}; \node[vtx] at (V) {};
    \node[bg] (B) at (1.3,-1.55) {}; \draw[gluon] (B)--(V);
    \node at (1.3,-2.05) {(b1)};
  \end{scope}
  \begin{scope}[shift={(6.8,0)}]
    \draw[quark] (0,0)--(0.55,0); \draw[quark] (0.55,0)--(2.05,0); \draw[quark] (2.05,0)--(2.6,0);
    \coordinate (V) at (1.3,-0.75);
    \draw[gluon] (0.55,0)--(V);
    \draw[gluon] (2.05,0)--($(2.05,0)!0.5!(V)$); \draw[aux] ($(2.05,0)!0.5!(V)$)--(V);
    \node[vtx] at (0.55,0) {}; \node[vtx] at (2.05,0) {}; \node[vtx] at (V) {};
    \node[bg] (B) at (1.3,-1.55) {}; \draw[gluon] (B)--(V);
    \node at (1.3,-2.05) {(b1$'$)};
  \end{scope}
  \begin{scope}[shift={(10.2,0)}]
    \draw[quark] (0,0)--(0.55,0); \draw[quark] (0.55,0)--(2.05,0); \draw[quark] (2.05,0)--(2.6,0);
    \coordinate (V) at (1.3,-0.75);
    \draw[gluon] (0.55,0)--(V); \draw[gluon] (2.05,0)--(V);
    \node[vtx] at (0.55,0) {}; \node[vtx] at (2.05,0) {}; \node[mag] at (V) {};
    \node[bg] (B) at (1.3,-1.55) {}; \draw[gluon] (B)--(V);
    \node at (1.3,-2.05) {(b2)};
  \end{scope}
  \begin{scope}[shift={(5.1,-2.9)}]
    \draw[quark] (0,0)--(0.55,0); \draw[quark] (0.55,0)--(2.05,0); \draw[quark] (2.05,0)--(2.6,0);
    \coordinate (V) at (1.3,-0.75);
    \draw[gluon] (0.55,0)--(V); \draw[gluon] (2.05,0)--(V);
    \node[vtx] at (0.55,0) {}; \node[vtx] at (2.05,0) {}; \node[gfv] at (V) {};
    \node[bg] (B) at (1.3,-1.55) {}; \draw[gluon] (B)--(V);
    \node at (1.3,-2.05) {(b3)};
  \end{scope}
\end{tikzpicture}
\caption{One-loop corrections to the background quark--gluon vertex in the first-order formalism. (a): gluon exchange with the background field on the quark line. (b1), (b1$'$): GBQ vertex with the auxiliary leg (double line) closed on a mixed propagator. (b2): magnetic vertex (square). (b3): background gauge-fixing vertex (open circle). The three graphs (b1)--(b3) replace the single triple-gluon graph of the second-order formalism.}
\label{fig:vertex}
\end{figure}\begin{figure}[t]
\centering
\begin{tikzpicture}[scale=0.78]
\newcommand{\qv}[3]{
 \begin{scope}[shift={(#1,0)}]
  \draw[quark] (0,0)--(0.45,0); \draw[quark] (0.45,0)--(1.75,0); \draw[quark] (1.75,0)--(2.2,0);
  \coordinate (A) at (0.45,0); \coordinate (C) at (1.75,0); \coordinate (V) at (1.1,-0.75);
  #2
  \node[vtx] at (A) {}; \node[vtx] at (C) {}; \node[#3] at (V) {};
  \node[bg] (B) at (1.1,-1.5) {}; \draw[gluon] (B)--(V);
 \end{scope}}
 \qv{0}{\draw[gluon] (A)--(V); \draw[gluon] (C)--(V);}{blob}
 \node at (2.65,-0.5) {$=$};
 \qv{3.1}{\draw[gluon] (A)--($(A)!0.5!(V)$); \draw[aux] ($(A)!0.5!(V)$)--(V); \draw[gluon] (C)--(V);}{vtx}
 \node at (5.75,-0.5) {$+$};
 \qv{6.2}{\draw[gluon] (A)--(V); \draw[gluon] (C)--($(C)!0.5!(V)$); \draw[aux] ($(C)!0.5!(V)$)--(V);}{vtx}
 \node at (8.85,-0.5) {$+$};
 \qv{9.3}{\draw[gluon] (A)--(V); \draw[gluon] (C)--(V);}{mag}
 \node at (11.95,-0.5) {$+$};
 \qv{12.4}{\draw[gluon] (A)--(V); \draw[gluon] (C)--(V);}{gfv}
 \node at (1.1,-2.15) {\small $\frac12\xi C_A$};
 \node at (4.2,-2.15) {\small $\frac38(1+\xi)C_A$};
 \node at (7.3,-2.15) {\small $\frac38(1+\xi)C_A$};
 \node at (10.4,-2.15) {\small $0$};
 \node at (13.5,-2.15) {\small $-\frac14(3+\xi)C_A$};
\end{tikzpicture}
\caption{The non-Abelian part of the background quark--gluon vertex. The single second-order graph with Abbott's background vertex (blob) equals the sum of the four first-order graphs of Fig.~\ref{fig:vertex}. Below each graph is the coefficient of its ultraviolet pole in units of $\K\gamma_\lambda$, Eq.~\eqref{eq:Lambda}.}
\label{fig:vertexsum}
\end{figure}
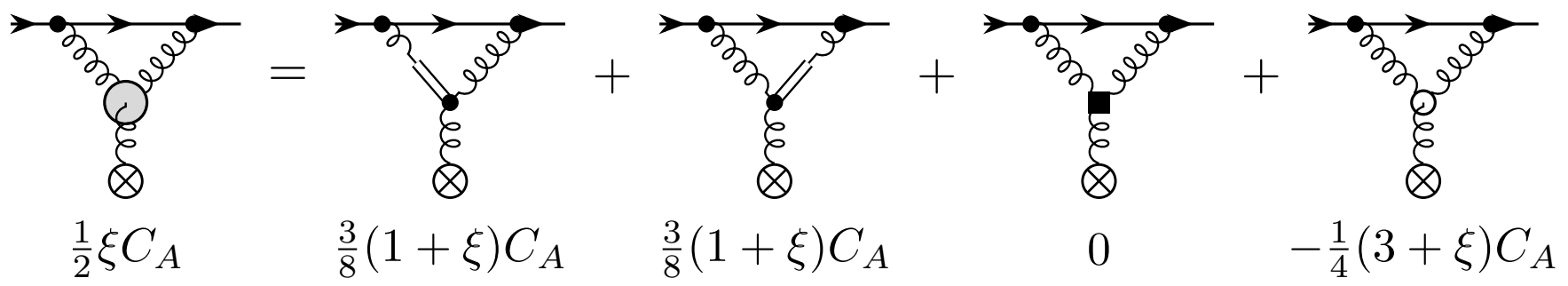\section{Gluonic part of the background two-point function}
\label{sec:twopoint}

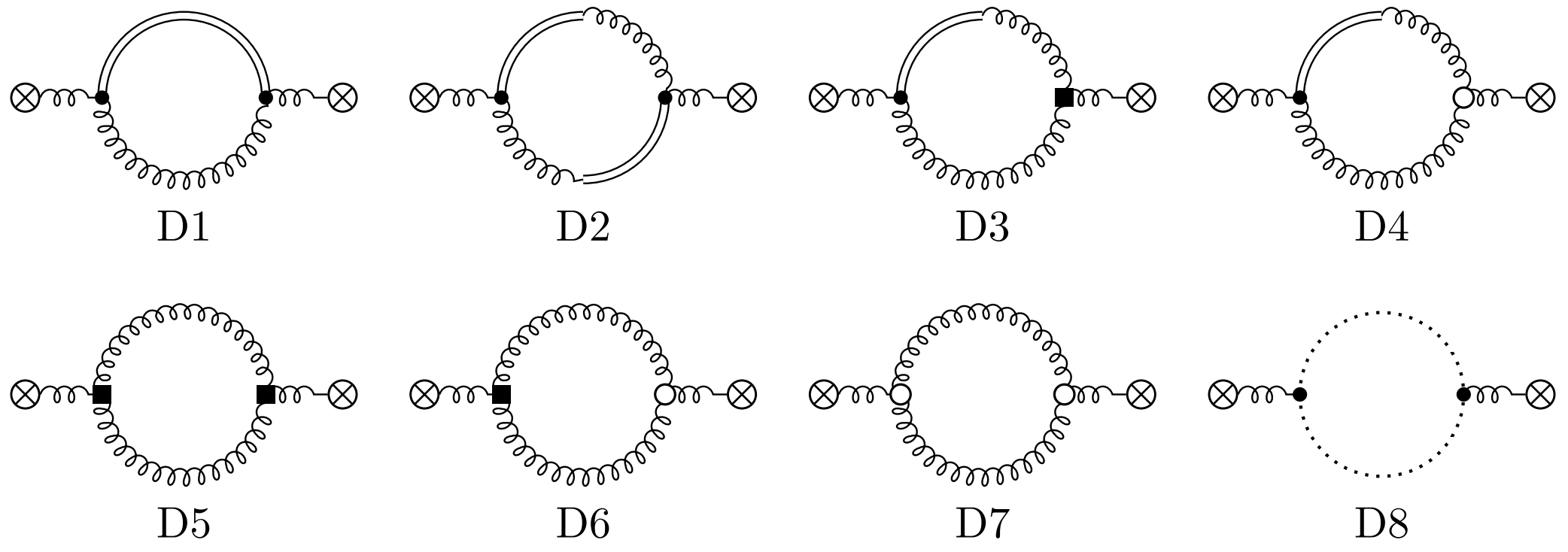
\begin{figure}[t]
\centering
\newcommand{\toposkel}[4]{
  \node[bg] (BL) at (-0.75,0) {}; \node[bg] (BR) at (2.35,0) {};
  \coordinate (L) at (0,0); \coordinate (R) at (1.6,0);
  \draw[gluon] (BL)--(L); \draw[gluon] (R)--(BR);
  #3 #4
  \node[#1] at (L) {}; \node[#2] at (R) {};}
\newcommand{\up}[1]{\draw[#1] (0,0) arc[start angle=180,end angle=0,radius=0.8];}
\newcommand{\dn}[1]{\draw[#1] (0,0) arc[start angle=180,end angle=360,radius=0.8];}
\newcommand{\upmix}{\draw[aux] (0,0) arc[start angle=180,end angle=90,radius=0.8]; \draw[gluon] (0.8,0.8) arc[start angle=90,end angle=0,radius=0.8];}
\newcommand{\dnmixR}{\draw[gluon] (0,0) arc[start angle=180,end angle=270,radius=0.8]; \draw[aux] (0.8,-0.8) arc[start angle=270,end angle=360,radius=0.8];}
\begin{tikzpicture}[scale=0.9]
  \begin{scope}[shift={(0,0)}]   \toposkel{vtx}{vtx}{\up{aux}}{\dn{gluon}} \node at (0.8,-1.25) {D1}; \end{scope}
  \begin{scope}[shift={(3.9,0)}] \toposkel{vtx}{vtx}{\upmix}{\dnmixR} \node at (0.8,-1.25) {D2}; \end{scope}
  \begin{scope}[shift={(7.8,0)}] \toposkel{vtx}{mag}{\upmix}{\dn{gluon}} \node at (0.8,-1.25) {D3}; \end{scope}
  \begin{scope}[shift={(11.7,0)}] \toposkel{vtx}{gfv}{\upmix}{\dn{gluon}} \node at (0.8,-1.25) {D4}; \end{scope}
  \begin{scope}[shift={(0,-2.9)}]   \toposkel{mag}{mag}{\up{gluon}}{\dn{gluon}} \node at (0.8,-1.25) {D5}; \end{scope}
  \begin{scope}[shift={(3.9,-2.9)}] \toposkel{mag}{gfv}{\up{gluon}}{\dn{gluon}} \node at (0.8,-1.25) {D6}; \end{scope}
  \begin{scope}[shift={(7.8,-2.9)}] \toposkel{gfv}{gfv}{\up{gluon}}{\dn{gluon}} \node at (0.8,-1.25) {D7}; \end{scope}
  \begin{scope}[shift={(11.7,-2.9)}] \toposkel{vtx}{vtx}{\up{ghost}}{\dn{ghost}} \node at (0.8,-1.25) {D8}; \end{scope}
\end{tikzpicture}
\caption{Gluonic one-loop contributions to the background two-point function in the first-order formalism. Double lines are auxiliary-field propagators, half-double lines mixed propagators, squares magnetic vertices, and open circles gauge-fixing vertices. D3, D4 and D6 occur twice, with the two background legs exchanged. D8 is the ghost loop. Seagull graphs are scaleless and vanish.}
\label{fig:topologies}
\end{figure}

The quark part of $Z_B$ was obtained in Section~\ref{sec:mixing}. Here we give the gluonic part from the first-order rules. Every background vertex with two quantum legs has colour structure $f^{abc}$, so we write the part of the Lagrangian linear in $B$ as $B^a_\lambda J^{a\lambda}$ with $J^{a\lambda}=f^{acd}\Phi^c_iA^\lambda_{ij}\Phi^d_j$, where $\Phi_i$ runs over $Q_\rho$ and $G_{\mu\nu}$. Wick's theorem then gives the two-point vertex as a direct and a crossed contraction,
\begin{equation}
i\Pi^{ab\,\lambda\kappa}(k)=-C_A\delta^{ab}\!\int\!\frac{\dd^d\ell}{(2\pi)^d}\Big[A^\lambda_{ij}A^\kappa_{kl}\,\Delta_{ik}(\ell)\Delta_{jl}(-\ell-k)-A^\lambda_{ij}A^\kappa_{kl}\,\Delta_{il}(\ell)\Delta_{jk}(-\ell-k)\Big],
\label{eq:master}
\end{equation}
where $\Delta_{ij}$ is the matrix of propagators \eqref{eq:DQQ}--\eqref{eq:DGG}. The resulting topologies are shown in Fig.~\ref{fig:topologies}. D1 and D2 come from two GBQ vertices, closed by $\vev{GG}\vev{QQ}$ or by two mixed propagators; D3, D4 and D6 combine two different couplings and occur twice; D5 and D7 contain two magnetic or two gauge-fixing vertices; D8 is the ghost loop. The local part of $\vev{GG}$ in D1 multiplies a scaleless integral and vanishes, which is the first-order counterpart of the vanishing second-order seagull tadpole.

The poles follow from Eq.~\eqref{eq:UVproj} and the three integrals, in units of $i/16\pi^2\eb$ and over the measure $[\ell^2(\ell+k)^2]^{-1}$,
\begin{equation}
\mathcal P[1]=1,\qquad \mathcal P[\ell^\alpha]=-\frac12k^\alpha,\qquad \mathcal P[\ell^\alpha\ell^\beta]=\frac13k^\alpha k^\beta-\frac1{12}k^2\eta^{\alpha\beta}.
\label{eq:Pints}
\end{equation}
For example, in Feynman gauge D1 reduces to $C_Ag^2\int(2\ell^\lambda\ell^\kappa+\ell^2\eta^{\lambda\kappa})/[\ell^2(\ell+k)^2]$, which gives $(a,b)_{\rm D1}=(\frac23,-\frac16)$ in the notation of Table~\ref{tab:topologies}; the other topologies follow in the same way. Only $\vev{QQ}$ depends on $\xi$, and a $\vev{QQ}$ line ending on the derivative leg of a gauge-fixing vertex loses its $\xi$ dependence, so D2, D4, D7 and D8 are gauge independent. The longitudinal parts of D1, D3, D5 and D6 give the general-$\xi$ entries of the table; their $\xi$-dependent parts, $\frac12\xi(1,-1)$ from D1 and D3 and $-\frac12\xi(1,-1)$ from D5 and D6, cancel in the sum. The total,
\begin{equation}
\Pi^{ab}_{\lambda\kappa}(k)\Big|_{\rm YM,UV}=-\frac{11}3C_A\K\,\delta^{ab}\left(k_\lambda k_\kappa-k^2\eta_{\lambda\kappa}\right),
\label{eq:PiYM}
\end{equation}
is gauge independent and transverse, although no single topology is both. Figure~\ref{fig:secondorder} shows how the second-order graphs map onto the first-order topologies. Every topology with a magnetic vertex is transverse by itself, and the magnetic sector gives $-\frac12(5+\xi)$, which is three quarters of the paramagnetic part $-4$ of $-\frac{11}3=-4+\frac13$ \cite{Nielsen:1981,Hughes:1980} in Feynman gauge. Adding the quark pole \eqref{eq:Piqpole} and the counterterm $-\frac14\delta Z_Bf(B)\cdot f(B)$ gives $Z_B=1+(\frac{11}3C_A-\frac43T_FN_f)\K$, as used in Section~\ref{sec:mixing}.

\begin{table}[t]
\centering
\renewcommand{\arraystretch}{1.35}
\begin{tabular}{lcccc}
\toprule
 & \multicolumn{2}{c}{Feynman gauge} & \multicolumn{2}{c}{general covariant gauge $\xi$}\\
Topology & $a_i$ & $b_i$ & $a_i$ & $b_i$\\
\midrule
D1 (GBQ--GBQ, $\vev{GG}\vev{QQ}$) & $\frac23$ & $-\frac16$ & $\frac16(1+3\xi)$ & $\frac16(2-3\xi)$\\
D2 (GBQ--GBQ, $\vev{GQ}\vev{QG}$) & $-\frac12$ & $-\frac14$ & $-\frac12$ & $-\frac14$\\
D3 (GBQ--mag, $\times2$) & $-1$ & $1$ & $-\frac12(3-\xi)$ & $\frac12(3-\xi)$\\
D4 (GBQ--gf, $\times2$) & $-\frac13$ & $\frac56$ & $-\frac13$ & $\frac56$\\
D5 (mag--mag) & $-1$ & $1$ & $-\frac12(1+\xi)$ & $\frac12(1+\xi)$\\
D6 (mag--gf, $\times2$) & $-1$ & $1$ & $-\frac12(1+\xi)$ & $\frac12(1+\xi)$\\
D7 (gf--gf) & $-\frac16$ & $-\frac1{12}$ & $-\frac16$ & $-\frac1{12}$\\
D8 (ghost loop) & $-\frac13$ & $\frac13$ & $-\frac13$ & $\frac13$\\
\midrule
Sum & $-\frac{11}3$ & $\frac{11}3$ & $-\frac{11}3$ & $\frac{11}3$\\
\bottomrule
\end{tabular}
\caption{Ultraviolet poles of the gluonic topologies of Fig.~\ref{fig:topologies}, in units of $C_A\K\delta^{ab}$, written as $a_ik^\lambda k^\kappa+b_ik^2\eta^{\lambda\kappa}$. Entries marked $\times2$ include both assignments of the two background legs.}
\label{tab:topologies}
\end{table}
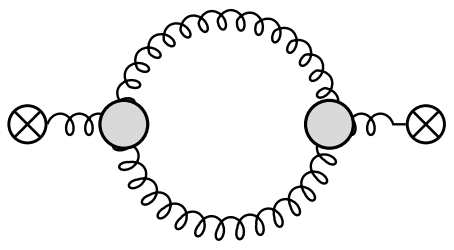
\begin{figure}[t]
\centering
\begin{tikzpicture}[scale=0.85]
 \begin{scope}
  \node[bg] (BL) at (-0.75,0) {}; \node[bg] (BR) at (2.35,0) {};
  \draw[gluon] (BL)--(0,0); \draw[gluon] (1.6,0)--(BR);
  \draw[gluon] (0,0) arc[start angle=180,end angle=0,radius=0.8];
  \draw[gluon] (0,0) arc[start angle=180,end angle=360,radius=0.8];
  \node[blob] at (0,0) {}; \node[blob] at (1.6,0) {};
  \node[right] at (2.8,0) {$=\ \ {\rm D1}+{\rm D2}+{\rm D3}+{\rm D4}+{\rm D5}+{\rm D6}+{\rm D7}\Big|_{\rm nonlocal}$};
 \end{scope}
 \begin{scope}[shift={(0,-2.3)}]
  \coordinate (V) at (0.8,-0.2);
  \node[bg] (BL) at (-0.1,-0.2) {}; \node[bg] (BR) at (1.7,-0.2) {};
  \draw[gluon] (BL)--(V); \draw[gluon] (V)--(BR);
  \draw[gluon] (V) arc[start angle=270,end angle=-90,radius=0.55];
  \node[vtx] at (V) {};
  \node[right] at (2.8,0) {$=\ \ {\rm D1}\Big|_{\rm local}\ =\ 0$\quad (scaleless)};
 \end{scope}
 \begin{scope}[shift={(0,-4.4)}]
  \node[bg] (BL) at (-0.75,0) {}; \node[bg] (BR) at (2.35,0) {};
  \draw[gluon] (BL)--(0,0); \draw[gluon] (1.6,0)--(BR);
  \draw[ghost] (0,0) arc[start angle=180,end angle=0,radius=0.8];
  \draw[ghost] (0,0) arc[start angle=180,end angle=360,radius=0.8];
  \node[vtx] at (0,0) {}; \node[vtx] at (1.6,0) {};
  \node[right] at (2.8,0) {$=\ \ {\rm D8}$};
 \end{scope}
\end{tikzpicture}
\caption{Second-order background self-energy graphs and their first-order counterparts. The gluon loop with two Abbott vertices (blobs) splits into the seven first-order topologies D1--D7 of Fig.~\ref{fig:topologies}; the second-order seagull tadpole corresponds to the local part of $\vev{GG}$ in D1, and both vanish in dimensional regularization; the ghost loop is unchanged.}
\label{fig:secondorder}
\end{figure}
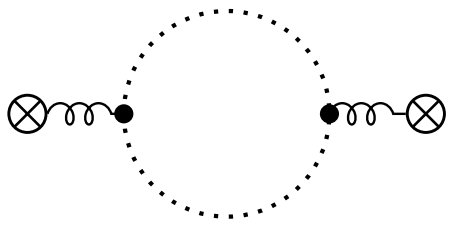\section{Stationarity argument and the bilinear identity}
\label{app:stationarity}

Let $\Lag(F)=\Lag_0(F)+\varepsilon\Lag_1(F)$, where $\Lag_0$ is stationary at $F_0$ and $\varepsilon=O(1/M^2)$. Eliminating $F$ gives $F=F_0+\varepsilon F_1+O(\varepsilon^2)$ and
\begin{equation}
\Lag\Big|_{\rm sol}=\Lag_0(F_0)+\varepsilon\left[\Lag_1(F_0)+F_1\cdot\frac{\partial\Lag_0}{\partial F}\Big|_{F_0}\right]+O(\varepsilon^2)=\Lag_0(F_0)+\varepsilon\Lag_1(F_0)+O(\varepsilon^2),
\label{eq:stationarity}
\end{equation}
because $\partial\Lag_0/\partial F$ vanishes at $F_0$. Two choices of $\Lag_1$ that agree at $F_0=f(A)$ therefore give the same Lagrangian for $A$ up to $O(1/M^4)$. This justifies replacing $\mathcal O_i$ by $\mathcal O_i^F$ in Section~\ref{sec:threshold}. Applied to $\frac14F\cdot F-\frac12F\cdot f+(a_h/M^2)F\cdot D^2F$, it gives back $-\frac14f\cdot f+(a_h/M^2)f\cdot D^2f$, Eq.~\eqref{eq:a}.

\paragraph{The bilinear identity.} For a plane wave $A_\mu e^{-ik\cdot x}$, $f_{\mu\nu}=-i(k_\mu A_\nu-k_\nu A_\mu)$, so $f^*\cdot\partial^2f=-k^2\,(k\wedge A)^*\cdot(k\wedge A)=-2k^2\left(k^2|A|^2-|k\cdot A|^2\right)$, while $\partial^\mu f_{\mu\nu}=-(k^2A_\nu-k_\nu\,k\cdot A)$ gives $(\partial f)^*\cdot(\partial f)=k^2\left(k^2|A|^2-|k\cdot A|^2\right)$. Hence $f\cdot\partial^2f=-2(\partial^\mu f_{\mu\nu})(\partial_\rho f^{\rho\nu})$ at the bilinear level, which fixes $C_1=-2a_h$ in Eq.~\eqref{eq:C1}.

\bibliographystyle{JHEP}
\bibliography{refs}

\end{document}